# INTERACTING FERMIONS IN ONE DIMENSION: FROM WEAK TO STRONG CORRELATION

H.J. SCHULZ
Laboratoire de Physique des Solides
Université Paris–Sud, 91405 Orsay, France

## 1 Introduction

A theoretical understanding of interacting fermion systems in one dimension is important for a number of reasons. On the one hand, in the physics of quasi-one-dimensional organic conductors [Jérome and Schulz, 1982] or of conducting polymers [Heeger et al., 1988] interaction effects play a major role. On the other hand, one–dimensional models can be easier to understand than their higher-dimensional versions, or even exactly solvable, as is the case with the prototypical model of correlated fermions, the Hubbard model [Lieb and Wu, 1968]. They therefore can provide valuable information on the role of correlation effects in higher dimension, e.g. on the physics of correlated fermions in two dimensions which is thought to be at the origin of the many interesting properties of high-temperature superconductors [Anderson, 1987, Anderson, 1988].

Theoretical work on interacting fermions in one dimension has progressed along a number of different lines. One approach has been the perturbative investigation of the weak coupling limit. Even this is in fact not entirely straightforward, mainly because of the infrared divergences encountered in this type of calculation which require a renormalization group treatment. An elementary introduction is given in chapter 2 below, and a very complete review can be found in the literature [Sólyom, 1979]. An alternative and more general approach is provided by the so–called "bosonization" method, which is based on the equivalence (valid in certain cases) between interacting fermions and noninteracting bosons and on the expression of fermionic operators in terms of bosons. Combined with the renormalization group approach, the bosonization method provides a rather straightforward description of the peculiar properties of one–dimensional interacting fermion systems ("Luttinger liquid"), and one finds that the low–energy physical properties are determined by only three parameters: the velocities of collective charge- and spin-density oscillations ($u_{\rho,\sigma}$), and a coefficient $K_\rho$ that determines the long–distance decay of correlation functions. These coefficients play a role similar to the Landau parameters of (three–dimensional) Fermi liquid theory. A number of physical properties depending on these parameters are discussed below, but let us mention here that in particular the coefficient $K_\rho$ is important in a much wider variety of phenomena: the temperature dependence of the NMR relaxation rate [Bourbonnais et al., 1984] or of X-ray scattering intensities [Pouget, 1988], effect of impurities [Giamarchi and Schulz, 1988], or possible low-temperature ordered states in systems of coupled chains all depend on it. A brief discussion of bosonization will be given in chapter 3, but for more detailed and rigorous derivations and results, the reader is referred to more specialized articles [Emery, 1979, Emery, 1992].

A rather different approach (at least until recently) is based on the famous "Bethe ansatz" [Bethe, 1931] which in particular has made possible an exact solution of both continuum fermions interacting via $\delta$–function potentials [Gaudin, 1967, Yang, 1967] and of the one–dimensional Hubbard model [Lieb and Wu, 1968] (and of many other interesting models). Without going into the mathematical details, in section 4.2 I will try to explain the basic ingredients of the exact solution of the Hubbard model, and then discuss in some detail the resulting spectrum of low–lying excitations. This will give a rather concrete illustration of the concept of "holons" and "spinons".

The exact Bethe ansatz eigenfunctions are so complicated that the direct calculation of correlation functions like (3.17), (3.18) and other physical properties of the one–dimensional Hubbard model is hard even for very small systems [Ogata and Shiba, 1990] and impossible in the thermodynamic limit. In section 4.3 I present a method that allows in particular a determination of the coefficient $K_\rho$ for arbitrary correlation strength. One then obtains a rather detailed *and exact* description of the low–energy (and low temperature) properties and also of the metal–insulator transition occurring when the average particle number per site, $n$, approaches unity. The method generalizes rather straightforwardly to other models.

Chapters 2 and 3 are intended to be brief and, hopefully, pedagogical introductions to subjects where detailed reviews exist, as cited above. On the other hand, the material in chapter 4 is to a large extent rather recent, and in some cases previously unpublished.

# 2  Weak Coupling Case

## 2.1  The Model

In this chapter we are interested in weakly interacting fermions in a one–dimensional metal. I consider a simple one–band model. The "kinetic energy" term in the Hamiltonian (which really also contains the interaction with the static crystal potential) then is of the form

$$H_0 = \sum_{k,s} \varepsilon_k c^\dagger_{k,s} c_{k,s} \quad , \tag{2.1}$$

where $c_{k,s}$ and $c^\dagger_{k,s}$ are the standard annihilation and creation operators for a fermion with momentum $k$ and spin s. $\varepsilon_k$ is the single–particle bandstructure. In a simple tight–binding model one would have $\varepsilon_k = -2t\cos k$ (the lattice constant is set to unity), but the precise form of $\varepsilon_k$ is unimportant in this chapter. The Fermi surface consists just of the two points $\pm k_F$.

For weak interactions between the particles, only states in the immediate vicinity of the Fermi points are important. For these states, one then can linearize the electronic dispersion relation around the Fermi points, and the kinetic energy term takes the form

$$H_0 = v_F \sum_{k,s} \{(k - k_F)a^\dagger_{k,s}a_{k,s} + (-k - k_F)b^\dagger_{k,s}b_{k,s}\} \;, \tag{2.2}$$

Here the $a$ ($b$) operators refer to states in the vicinity of $+k_F$ ($-k_F$), i.e. the $a$–particles move to the right, the $b$–particles move to the left. The $k$–summation is limited to an interval $[-k_0, k_0]$ around $k_F$ (typically, $k_0 \approx \pi$, but the precise value isn't important here). The Fermi velocity is given by

$$v_F = \left.\frac{\partial \varepsilon_k}{\partial k}\right|_{k_F} \;, \tag{2.3}$$

and the density of states per spin is $N(E_F) = 1/(\pi v_F)$.

We now introduce interactions between the fermions. For simplicity, I will only consider here the so–called forward $(k_F, s; -k_F, t) \to (k_F, s; -k_F, t)$ and backward $(k_F, s; -k_F, t) \to (-k_F, s; k_F, t)$ scattering processes, with coupling constants $g_2$ and $g_1$, respectively. The operator decribing these processes is

$$H_{int} = \frac{1}{L} \sum_{kpqst} [g_1 a^\dagger_{k,s} b^\dagger_{p,t} a_{p+2k_F+q,t} b_{k-2k_F-q,s} + g_2 a^\dagger_{k,s} b^\dagger_{p,t} b_{p+q,t} a_{k-q,s}] \;. \tag{2.4}$$

Again, only processes involving particles close to the Fermi surface are important, and therefore the dependence of the scattering potentials $g_{1,2}$ on the momentum transfer $q$ (and possibly on the other momenta) is neglected. For Coulomb interactions one expects $g_1, g_2 > 0$. In principle, the long–range part of the Coulomb repulsion leads to a singular $q$–dependence. Such singularities in $g_2$ can be handled rather straightforwardly by the techniques discussed in chapter 2, but here I shall limit myself to constants $g_1, g_2$. Electron–phonon interactions can lead to effectively attractive interactions between electrons, and therefore in the following I will not make any restrictive assumptions about the sign of the constants. One should however notice that a proper treatment of the phonon dynamics and of the resulting retardation effects requires more care [Voit and Schulz, 1988]. The model thus defined (only forward and backward scattering, *constants* $g_1$ and $g_2$) exhibits already both the fundamental physical phenomena of one–dimensional fermions systems, and the technical problems encountered in its solutions, and I will therefore concentrate on this model for the rest of this chapter.

In order to be more realistic, one can to introduce other types of interaction: on the one hand, in a half–filled band, umklapp scattering $(k_F, s; k_F, t) \rightarrow (-k_F, s; -k_F, t)$ is possible, because the total change in momentum then is a reciprocal lattice vector. On the other hand, there is also the so–called $g_4$ process $(k_F, s; k_F, t) \rightarrow (k_F, s; k_F, t)$. These types of interaction can lead to new phenomena in some cases, but the methods used for their treatment are similar to the ones discussed below for the $g_1 - g_2$ model.

## 2.2 Mean–Field Theory (beware!)

The model introduced above exhibits various kinds of instabilities. A qualitative idea about this can be obtained using mean–field theory. It should however be said right from the outset that this is going to produce a number of qualitatively wrong results insofar as ground states with a broken symmetry are found, whereas exact theorems forbid symmetry breaking.

To demonstrate the occurrence of instabilities in the framework of a mean–field theory, let us in particular consider the so–called Peierls instability, e.g. the instability of an interacting one–dimensional fermion gas against the formation of a charge density wave with wavevector $2k_F$. The corresponding order parameter is

$$\Delta_{CDW} = \langle \sum_{k,s} b^\dagger_{k,s} a_{k+2k_F,s} \rangle \ . \tag{2.5}$$

Note that $\Delta_{CDW}$ is a complex quantity. Introducing the operator for fluctuations around the mean–field expectation value

$$\delta_q = \sum_{k,s} b^\dagger_{k,s} a_{k+2k_F+q,s} - \langle \sum_{k,s} b^\dagger_{k,s} a_{k+2k_F+q,s} \rangle \ , \tag{2.6}$$

and neglecting all terms of order $\delta^2$, one finds

$$H_{int} \approx \Delta_{CDW} \sum_{k,s} a^\dagger_{k+2k_F,s} b_{k,s} + \Delta^*_{CDW} \sum_{k,s} b^\dagger_{k,s} a_{k+2k_F,s} \ . \tag{2.7}$$

The approximate Hamiltonian then is in bilinear form, and one straightforwardly finds the energy eigenvalues (for $k > 0$)

$$E_k = \text{sign}(k - k_F)\sqrt{v_F^2(k - k_F)^2 + |\Delta_{CDW}|^2} \ , \tag{2.8}$$

and a similar expression for $k < 0$. One now sees that the energy of all occupied states is *lowered* if $\Delta_{CDW} \neq 0$. The total energy gain is easily calculated (for small $\Delta_{CDW}$) as

$$\Delta E \approx |\Delta_{CDW}|^2 \ln(|\Delta_{CDW}|/E_0) \ , \tag{2.9}$$

and because of the logarithmic factor, it is always favorable to have $|\Delta_{CDW}| \neq 0$.

The most important properties of the state with $|\Delta_{CDW}| \neq 0$ are:

1. The average charge density is modulated, i.e. one has a charge–density wave (CDW):

$$\langle \rho(x) \rangle = \rho_0 + \delta\rho \cos(2k_F x + \phi) \ , \tag{2.10}$$

where $\phi$ is the phase of $\Delta_{CDW}$, and $\rho_0$ is the average charge density. The CDW thus break,s the continuous translational symmetry of the $g_1 - g_2$ model, variations of $\phi$ representing continuous translations of the CDW. The breaking of the continuous symmetry implies that there are low–energy collective excitations (Goldstone modes) with a linear energy–momentum relation $\omega(k) = c|k|$, with some velocity $c$. These excitations represent long–wavelength fluctuations of the wavelength of the CDW.

2. There is a gap in the spectrum of single particle excitations. Consequently, there is no Pauli susceptibility, and no single–particle conductivity. However, the CDW can move as a whole, leading to the charge–density wave conductivity first proposed by Fröhlich.

One can investigate other types of instability along similar lines. The relevant operators which can take on non–zero expectation values are:

1. spin density wave (SDW):

$$O_{SDW_\alpha} = \sum_{k,s,t} b^\dagger_{k,s} \sigma^\alpha_{s,t} a_{k+2k_F,t} \ , \tag{2.11}$$

2. singlet superconductivity (SS):

$$O_{SS} = \sum_{k,s} s b_{-k,-s} a_{k,s} \ , \tag{2.12}$$

3. triplet superconductivity (TS):

$$O_{TS_\alpha} = \sum_{k,s,t} s b_{-k,-s} \sigma^\alpha_{s,t} a_{k,t} \ , \tag{2.13}$$

where $\sigma^\alpha$ are the Pauli spin matrices ($\alpha = x, y, z$). These operators are Fourier transforms of the real space operators

$$\begin{aligned}
O_{CDW}(x) &= \sum_s \psi^\dagger_{-,s}(x) \psi_{+,s}(x) \ , \\
O_{SDW_\alpha}(x) &= \sum_{s,t} \psi^\dagger_{-,s}(x) \sigma^\alpha_{s,t} \psi_{+,t}(x) \ , \\
O_{SS}(x) &= \sum_s s \psi_{-,-s}(x) \psi_{+,s}(x) \ , \\
O_{TS_\alpha}(x) &= \sum_{s,t} s \psi_{-,-s}(x) \sigma^\alpha_{s,t} \psi_{+,t}(x) \ ,
\end{aligned} \tag{2.14}$$

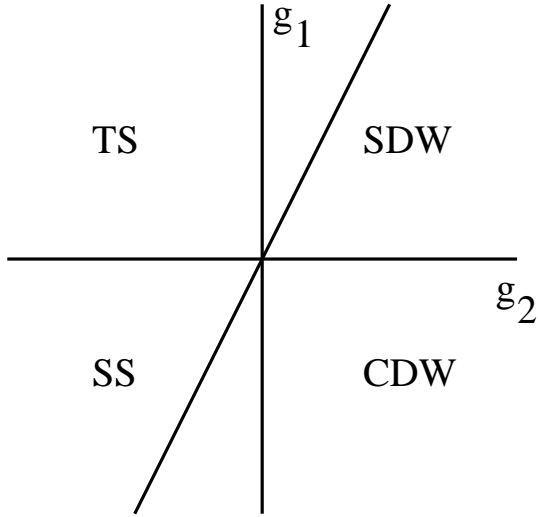

Figure 1: *Ground state phase diagram of the $g_1 - g_2$ model obtained in the mean–field approximation*

where $\psi_{r,s}(x)$ is the annihilation operator for a right–going ($r = +$) or left–going ($r = -$) particle with spin $s$.

Seeking the energetically most stable state in the $g_1 - g_2$–plane, one finds the "phase diagram" shown in fig. 1. All the states in fig. 1 break continuous symmetries: in the CDW and SDW states, translational symmetry is broken, in the SS and TS states gauge symmetry is broken. In addition, in the "triplet" phase SDW and TS, spin rotation invariance is also broken. Consequently, all theses states have at least one Goldstone mode in their excitation spectrum.

Compared to a three–dimensional interacting fermion system, fig. 1 exhibits a remarkable feature: in no part of the parameter space (except in the trivial singular point $g_1 = g_2 = 0$) is a "normal" (Fermi liquid like) state stable. While the pairing instabilities SS and TS are indeed the analogue of the well–known Cooper instability occuring in any dimension for weakly attractive forces between fermions, the density–wave type instabilities (CDW, SDW) are due to the particular form of the one–dimensional Fermi "surface", and have no direct analogues in higher dimensions (apart from special cases with nesting Fermi surfaces). In particular, weakly repulsive fermions in three dimensions are Fermi liquids without any signs of symmetry breaking.

One can extend the mean–field calculation in order to check its internal consistency. A particular useful check is a calculation of the fluctuations of the order parameter around its mean value. Without going into details, one finds

$$\langle |O_\gamma(x)|^2 \rangle - |\langle O_\gamma \rangle|^2 \approx \int_{|q|<1/\xi_0} d^d q \coth(\beta\omega_q/2)/\omega_q \quad . \tag{2.15}$$

Here $\omega_q = c|q|$ is the energy of the Goldstone modes, $\beta = 1/T$, $\gamma =$ CDW, SDW,

SS, or TS, and $\xi_0$ is the BCS coherence length for the pairing instabilities or an analogous length for the density wave instabilities. Expression (2.15) is divergent in one dimension both at zero and at finite temperature (and in two dimensions for $T \neq 0$), indicating that in these cases fluctuations of the order parameter about its supposed mean value are infinitely large. This clearly indicates that something went wrong in the mean–field theory, and that the assumption of a symmetry–broken ground state doesn't make sense. A more consistent treatment will be given in the following sections.

## 2.3 Renormalization Group: Coupling Constants

In order to understand the problems with the mean–field approach discussed at the end of the last section, it is helpful to take a step back and to look at the usual microscopic justification of the instability criteria used. The typical case is the BCS theory of superconductivity (in three dimensions). Here the mean–field critical temperature corresponds to a divergence of the pair susceptibility

$$\Delta(T) = \int dr d\tau \langle T_\tau \psi_s^\dagger(x,\tau) \psi_{-s}^\dagger(x,\tau) \psi_s(0,0) \psi_{-s}(0,0) \rangle \tag{2.16}$$

calculated in the random phase approximation. One then has

$$\Delta(T) = \Delta_0(T)/(1 + \lambda \Delta_0(T)) \ , \tag{2.17}$$

where $\Delta_0(T)$ is the noninteracting pair susceptibility. The standard justification for the approximation (2.17) lies in the fact that $\Delta_0(T)$ is logarithmically divergent at low temperatures, and that the RPA takes into account the most divergent diagrams at each order in perturbation theory.

In one dimension, things are more complicated: in addition to the logarithmic divergence in $\Delta_0(T)$, the density–density correlation function at wavevector $2k_F$

$$\Pi_0(T) = \frac{1}{L} \sum_k \frac{f(\varepsilon_k) - f(\varepsilon_{k-2k_F})}{\varepsilon_k - \varepsilon_{k-2k_F}} \tag{2.18}$$

is also logarithmically divergent. In a diagrammatic language, $\Delta_0$ and $\Pi_0$ are given by the particle–particle and particle–hole diagram in fig. 2. Consequently, at each order in perturbation theory, a whole set of mixed particle–particle and particle–hole diagrams has to be summed, leading to the so–called parquet equations. In the present context, this has been done by Bychkov et al. in 1966 [Bychkov et al., 1966]. However, instead of discussing their method, I will here discuss a renormalization group approach which is based on Anderson's "poor man's scaling" of the Kondo problem (where similar mixed divergences occur) [Anderson, 1970, Anderson et al., 1970]. This approach has been adapted to the one–dimensional fermion problem by Sólyom [Sólyom, 1979], and I here closely follow his presentation.

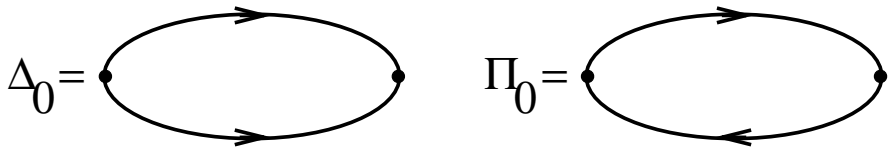

Figure 2: *The elementary logarithmically divergent particle–particle and particle–hole diagrams*

The basic idea is by now well known: the divergent terms are of the form $\ln(E_0/T)$ (or $\ln(E_0/\omega), \ln(E_0/v_F|q|)$ at $T=0$), where $E_0 = v_F k_0$. Consequently, if it is possible, via an elimination of high–energy degrees of freedom, to find a model with a smaller cutoff energy $E$, but the same physical properties at energies much lower than $E$, the problem of divergent diagrams is less serious. Moreover the problem may even be completely cured if it is possible to apply the transformation repeatedly until $E \approx T$ (provided one still remains within the range of applicability of perturbation theory).

Physical properties at low energies are determined by the two–particle T–matrix

$$T(\omega) = H_{int} + H_{int}\frac{1}{\omega + i\varepsilon - H_0}T \tag{2.19}$$

which is related to the transition probability between an initial state $|i\rangle$ and a final state $|f\rangle$ via

$$W_{i\to f} = 2\pi |\langle i|T(E_i)|f\rangle|^2 \rho(f) , \tag{2.20}$$

where $E_i$ is the energy of the initial (and final) state, and $\rho(f)$ is the density of final states. Consider initial and final states

$$|i\rangle = a^\dagger_{k_1,\alpha} b^\dagger_{k_2,\beta}|0\rangle \ , |f\rangle = b^\dagger_{k_3,\gamma} a^\dagger_{k_4,\delta}|0\rangle \ , \tag{2.21}$$

where $|0\rangle$ is the filled Fermi sea, and $k_1, k_4 \approx k_F, k_2, k_3 \approx -k_F$. Second order perturbation theory in $H_{int}$ gives

$$\langle i|T(\omega)|f\rangle = \frac{1}{L}\left\{\left(g_1 - \frac{g_1^2}{\pi v_F}\int_\omega^{E_0}\frac{dE}{E}\right)\delta_{\alpha,\gamma}\delta_{\beta,\delta} \right.$$
$$\left. - \left(g_2 - \frac{g_1^2}{2\pi v_F}\int_\omega^{E_0}\frac{dE}{E}\right)\delta_{\alpha,\delta}\delta_{\beta,\gamma}\right\} , \tag{2.22}$$

where in the summation over intermediate states of energy $E$ only the contributions divergent for $\omega \to 0$ have been retained. Note that the only nonzero contributions come from the $g_1$–interactions, all singular contributions from $g_2$ cancel each other.

The perturbative result (2.22) can now be used to generate renormalisation group equations. We want to reduce the cutoff $E_0$ to a lower value $E_0'$, and simultaneously adjust the coupling constants $g_i$, with the requirement that the physics at low energies $\omega \ll E_0'$ remains unchanged, i.e. that

$$\frac{d}{dE_0'}\langle i|T(\omega)|f\rangle = 0 \ . \tag{2.23}$$

Using the explicit result (2.22) and changing the name of the upper cutoff $E'_0 \to E$), one finds the *renormalization group equations*

$$E\frac{d}{dE}g_1(E) = \frac{1}{\pi v_F}g_1^2(E) ,$$
$$E\frac{d}{dE}g_2(E) = \frac{1}{2\pi v_F}g_1^2(E) . \qquad (2.24)$$

The initial conditions for these equations are that when the "running" cutoff $E$ equals the initial cutoff $E_0$, the coupling constants should take their bare values: $g_i(E_0) = g_i$.

One can look at this derivation from a slightly different point of view: perturbatively, the $g_1$–like term in $\langle i|T(\omega)|f\rangle$ (i.e. the term proportional to $\delta_{\alpha,\gamma}\delta_{\beta,\delta}$) is

$$T_1(\omega) = g_1 + \frac{g_1^2}{\pi v_F}\ln(E_0/\omega) , \qquad (2.25)$$

which can be rewritten, in terms of a reduced cutoff energy

$$T_1(\omega) = \left[g_1 + \frac{g_1^2}{\pi v_F}\ln(E_0/E'_0)\right] + \frac{g_1^2}{\pi v_F}\ln(E'_0/\omega) . \qquad (2.26)$$

This last expression now can be interpreted as a perturbative result for a system where the initial coupling constant and cutoff $(g_1, E_0)$ are replaced by new parameters $(g'_1, E'_0)$, with

$$g'_1 = g_1 + \frac{g_1^2}{\pi v_F}\ln(E_0/E'_0) . \qquad (2.27)$$

Using an $E'_0$ infinitesimally smaller than $E_0$, and iterating the transformation $(g_1, E_0) \to (g'_1, E'_0)$, one recovers the renormalization group equations (2.24).

The renormalization group equations (2.24) show that there is no renormalization of the coupling constants if $g_1 = 0$. The general solution for $g_1$ is

$$g_1(E) = \frac{g_1}{1 + \frac{g_1}{\pi v_F}\ln(E_0/E)} , \qquad (2.28)$$

and $g_2(E)$ is obtained straightforwardly noting that the combination $g_1 - 2g_2$ is unrenormalized. This result now shows that there are two qualitatively different regimes: if initially $g_1 > 0$, one has $g_1(E) \to 0$ when the cutoff energy decreases, i.e. one renormalizes to a weak–coupling regime. One then expects that the perturbative approach used here should give reliable results for the low–energy properties of the model. On the other hand, if initially $g_1 < 0$, $g_1(E)$ diverges at some finite cutoff energy. Clearly, close to the divergence, the assumption of weak coupling underlying the present expansion breaks down. The physical properties in this regime can however be understood using the bosonization method introduced in the next chapter. The cases $g_1 > 0$ and $g_1 < 0$ correspond to the ferromagnetic and antiferromagnetic Kondo problems, respectively.

## 2.4 Renormalization Group: Correlation functions

Let us now use the above results for $g_1 > 0$, where everything is well–defined, to calculate correlation functions. For definiteness, let us consider the correlation function of charge–density fluctuations with wavevector around $2k_F$, as represented diagrammatically in fig. 3. In RPA, this function diverges at the mean–field critical temperature

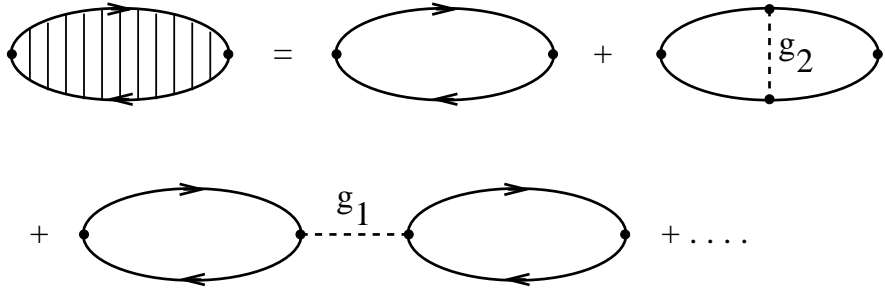

Figure 3: *Perturbativ expansion of the CDW correlation function $\mathcal{D}_{CDW}$ (the hatched bubble) to first order in the $g_i$.*

of the Peierls instability. To zeroth order one finds, at $T = 0$

$$\mathcal{D}^0_{CDW}(q,\omega) \equiv \mathcal{D}^0(q,\omega) = \frac{1}{\pi v_F} \ln \frac{[\omega^2 + v_F^2 q^2]^{1/2}}{E_0} \ . \tag{2.29}$$

The first order result is

$$\mathcal{D}_{CDW} = \mathcal{D}^0[1 + (g_{CDW}/2)\mathcal{D}^0 + ...] \ , \tag{2.30}$$

with $g_{CDW} = 2g_1 - g_2$. For the other types of instabilities (SDW, SS, TS), analogous results can be found, with coupling constants $g_{SDW} = -g_2, g_{SS} = g_1 + g_2, g_{TS} = -g_1 + g_2$.

We now use these perturbative results to obtain renormalization group equations for the correlation functions $\mathcal{D}_i$. The general structure of the perturbation theory is, at energy cutoff $E$

$$\mathcal{D}_i = \frac{1}{\pi v_F} \ln(\omega/E)[1 - \frac{g_i}{2\pi v_F} \ln(\omega/E) + ...] \ , \tag{2.31}$$

and therefore varying the cutoff, one has

$$\pi v_F d\mathcal{D}_i = [-1 + \frac{g_i}{\pi v_F} \ln(\omega/E)]\frac{dE}{E} \ . \tag{2.32}$$

We now have to take into account two effects: (i) if the running cutoff $E$ is smaller than the initial cutoff $E_0$, a certain number opf states have already been eliminated, and therefoe the coupling constant $g_i$ appearing in (2.32) actually is the renormalized coupling constant. (ii) for the same reason, the factor $1/(\pi v_F) \ln(\omega/E)$ appearing in (2.32), which in a purely perturbative calculation represents $\mathcal{D}^0$, should be replaced by the full correlation function at cutoff $E$. The renormalization group equation for $\mathcal{D}_i$ then becomes

$$\pi v_F \frac{d\mathcal{D}_i}{d\ln E} = -1 + g_i(E)\mathcal{D}_i(E) \ . \tag{2.33}$$

An initial condition is provided by the observation that when the cutoff is of the order of the external frequency $\omega$, the correlation function are finite and small, i.e. $\mathcal{D}_i(\omega; E \approx \omega) \approx 0$ (the precise value is of minor importance here). One then can integrate eq. (2.33), starting with the initial condition until the cutoff reaches $E_0$. The leading singular behaviour at low energies and momenta is found to be

$$\mathcal{D}_i(q,\omega) \approx [\omega^2 + v_F^2 q^2]^{-\alpha_i/2} \ln^{\nu_i}([\omega^2 + v_F^2 q^2]/E_0^2) \ . \tag{2.34}$$

Here the exponents $\alpha_i, \nu_i$ are

$$\begin{aligned}
\alpha_{CDW} &= \alpha_{SDW} &= (g_2 - g_1/2)/(\pi v_F) \ , \\
\alpha_{SS} &= \alpha_{TS} &= -(g_2 - g_1/2)/(\pi v_F) \ , \\
\nu_{SDW} &= \nu_{TS} &= 1/2 \ , \\
\nu_{CDW} &= \nu_{SS} &= -3/2 \ .
\end{aligned} \tag{2.35}$$

The important result here is the existence of power laws in the low–energy, low momentum regime, with the value of the exponents depending on the interaction constants, i.e. the exponents are non–universal. If $\alpha_i > 0$, the corresponding correlation function is enhanced over the logarithmic behaviour of the noninteracting case. Moreover, because of the logarithmic correction factors in (2.34), the SDW and TS correlations are always enhanced over the CDW and SS. The "phase diagram", indicating the strongest fluctuations in different parts of the $g_1 - g_2$ plane, is shown in fig. 4.

The real space correlation functions are found Fourier transforming (2.34) as

$$\mathcal{D}_i(r,\tau) \approx (r^2 + v_F^2\tau^2)^{-(1-\alpha_1/2)} \ln^{\nu_i}(r^2 + v_F^2\tau^2) \ , \tag{2.36}$$

i.e. as expected there is no long range order, even at zero temperature. However, provided that $\alpha_i > 0$, the spatial decay of correlations is slower than in the noninteracting case. Similary, at finite temperatures, one finds $\mathcal{D}_i(0,0) \approx (T/E_0)^{-\alpha_i} \ln^{\nu_i}(E_0/T)$, i.e. there are no singularities, and therefore no phase transitions at finite temperature, but the fluctuations with the strongest exponent $\alpha_i$ dominate at low temperature. Contrary to the correlation functions $\mathcal{D}_i$, thermodynamic properties like the spin susceptibility, the specific heat, or the compressibility have only finite corrections at low temperatures.

The calculations presented here are based on lowest nontrivial order perturbative calculations, and consequently, the exponents $\alpha_i$ are correct to order $g_i$. It is possible to go to higher order in the calculation. One then faces an interesting new phenomenon: self–energy corrections to the one–particle Green's function become singular at low energies and momenta (they are regular at first order). This signals the breakdown of the Landau quasiparticle picture in one dimension: in fact, as discussed in the next chapter, the elementary excitations at low energy are collective charge and spin fluctuations, rather than individual quasiparticles. As far as the renormalization group calculation is concerned, this means that it is no longer straightforward to define transition amplitudes between initial and final quasiparticle states as in (2.21, 2.22), because there are singular corrections to these states themselves. This situation can be handled by a field–theoretical formulation of the renormalization group, as described in detail by Sólyom [Sólyom, 1979]. This type of approach has been pursued up to third order in the coupling constants. One should however notice that in higher order, some of the integrations in the diagrammatic calculation are nonsingular, and therefore necessarily depend on details of the electronic bandstructure and on the cutoff procedure used. It is thus not very promising to push this approach to higher and higher order if one wants to understand strongly interacting fermions. On the other hand, the bosonization method allows one to obtain a number of general exact statements, which in some cases (e.g. the Hubbard model, see below) can be used to extract all the parameters determining the low energy physics, even in strongly interacting cases.

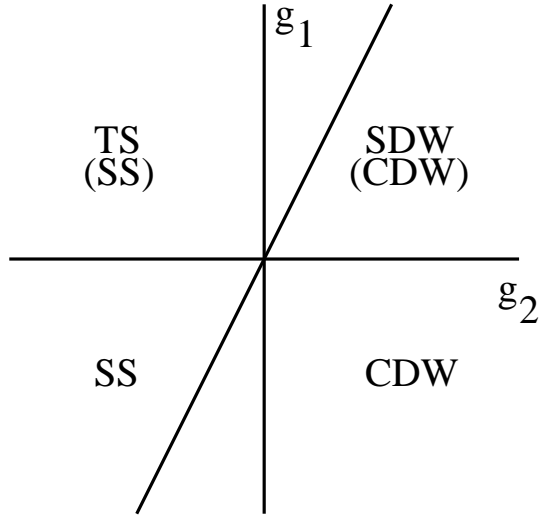

Figure 4: *Zero temperature "phase diagram" showing the most divergent type of flutuations in the $g_1-g_2$ plane. Phases in parentheses are logarithmically weaker divergent.*

# 3 Bosonization, spin–charge separation, Luttinger liquid

## 3.1 Bosonization Formalism

One of the important results in the theory of one–dimensional interacting fermions is that fermion field operators can be expressed in terms of boson operators [Luther and Peschel, 1974, Mattis, 1974]. This equivalence then can be used to express the fermion Hamiltonian in a particularly simple form, in terms of boson fields only (see eq.(3.3) below). Consider for the moment spinless electrons, and define a boson field $\phi(x)$ by $\partial \phi / \partial x = \pi \rho(x)$, where $\rho$ is the deviation from the average density. Then, creating a particle at site $x$ means introducing a kink of height $\pi$ in $\phi$, i.e. at points on the left of $x$ $\phi$ has to be shifted by $\pi$. Displacement operators are exponentials of momentum operators, and therefore a first guess would be $\psi^\dagger(x) \approx \exp(i\pi \int_{-\infty}^{x} \Pi(x') dx')$, where $\Pi$ is the momentum density conjugate conjugate to $\phi$: $[\phi(x), \Pi(y)] = i\delta(x-y)$. However, this operator commutes with itself, instead of satisfying canonical anticommutation relations. Anticommutation is achieved by multiplying with an operator, acting at site $x$, that changes sign each time a particle passes through $x$. Such an operator is $\exp(\pm i\phi(x))$. The final result then is

$$\psi^\dagger_\pm(x) = \lim_{\alpha \to 0} \frac{1}{\sqrt{2\pi\alpha}} \exp\left[\pm ik_F x \mp i\phi(x) + i\pi \int_{-\infty}^{x} \Pi(x') dx'\right] \;, \qquad (3.1)$$

where the upper and lower sign refer to electrons near $k_F$ and $-k_F$ respectively, and $\alpha$ is a short–distance cutoff. A detailed derivation of this important relation as an operator identity is given in the literature [Haldane, 1981a, Heidenreich et al., 1980]. One should also notice that (3.1) is strictly speaking valid for models with linear energy-momentum relation. If there is some curvature in the dispersion relation (as is necessarily the case in lattice models), all odd powers of $\exp(i\phi)$ appear [Haldane, 1981b], i.e.

$$\psi^\dagger_\pm(x) \approx \sum_{m \geq 0} \alpha_m \exp\left[\pm i(2m+1)(k_F x - \phi(x)) + i\pi \int_{-\infty}^{x} \Pi(x') dx'\right] \;, \qquad (3.2)$$

where $\alpha_m$ are model dependent constants.

For electrons with spin, one simply introduces one boson field for each spin orientation. Introducing charge and spin bosons via $\phi_{\rho,\sigma} = (\phi_\uparrow \pm \phi_\downarrow)/\sqrt{2}$, one then finds that the low-energy, large-distance behaviour of a one-dimensional fermion system with spin-independent interactions is described by the Hamiltonian [Sólyom, 1979, Emery, 1979]

$$H = H_\rho + H_\sigma + \frac{2g_1}{(2\pi\alpha)^2} \int dx \cos(\sqrt{8}\phi_\sigma) \;. \qquad (3.3)$$

Here $\alpha$ is a short-distance cutoff, $g_1$ is the backward scattering amplitude, and for $\nu = \rho, \sigma$

$$H_\nu = \int dx \left[ \frac{\pi u_\nu K_\nu}{2} \Pi_\nu^2 + \frac{u_\nu}{2\pi K_\nu} (\partial_x \phi_\nu)^2 \right] . \tag{3.4}$$

The full expressions for the phase fields are[*]

$$\phi_\nu(x) = -\frac{i\pi}{L} \sum_{p \neq 0} \frac{1}{p} e^{-\alpha|p|x/2 - ipx} [\nu_+(p) + \nu_-(p)] - N_\nu \frac{\pi x}{L} , \tag{3.5}$$

$$\Pi_\nu(x) = \frac{1}{L} \sum_{p \neq 0} e^{-\alpha|p|x/2 - ipx} [\nu_+(p) - \nu_-(p)] + J_\nu/L . \tag{3.6}$$

Here $\rho_r(p)$ ($\sigma_r(p)$) are the Fourier components of the charge (spin) density operator for right–($r = +$) and left–($r = -$) going fermions. Introducing the *total* number operators (measured with respect to the ground state) $N_{rs}$ for right– and left–going particles ($r = \pm$) of spin $s$, the (charge and spin) number and current operators are $N_\nu = [(N_{+\uparrow} + N_{-\uparrow}) \pm (N_{+\downarrow} + N_{-\downarrow})]/\sqrt{2}$, $J_\nu = [(N_{+\uparrow} - N_{-\uparrow}) \pm (N_{+\downarrow} - N_{-\downarrow})]/\sqrt{2}$, where the upper and lower sign refer to charge and spin, respectively.

The operators $\phi_\nu$ and $\Pi_\nu$ in (3.3) obey Bose–like commutation relations:

$$[\phi_\nu(x), \Pi_\mu(y)] = i \delta_{\nu\mu} \delta(x - y) ,$$

and consequently, at least for $g_1 = 0$, (3.3) describes independent long-wavelength oscillations of the charge and spin density, with linear dispersion relation $\omega_\nu(k) = u_\nu |k|$, and the system is conducting. This model (no backscattering), usually called the Tomonaga–Luttinger model, is the one to which the bosonization method was originally applied [Luttinger, 1963, Mattis and Lieb, 1965]. The only nontrivial interaction effects in (3.3) come from the cosine term. However, for repulsive interactions ($g_1 > 0$), this term is renormalized to zero in the long-wavelength limit, and at the fixed point one has $K_\sigma^* = 1$. The three remaining parameters in (3.3) then completely determine the long-distance and low–energy properties of the system. It should be emphasized that (3.3) can be derived exactly for fermions with linear energy–momentum relation. For more general (e.g. lattice) models, (3.3) is still the correct effective Hamiltonian for low–energy excitations.

The Hamiltonian (3.3) also provides an explanation for the physics of the case of negative $g_1$, where the renormalization group scales to strong coupling (cf. sec.2.3). In fact, if $|g_1|$ is large in (3.3), it is quite clear that the elementary excitations of $\phi_\sigma$ will be small oscillations around one of the minima of the cos term, or possibly soliton–like objects where $\phi_\sigma$ goes from one of the minima to the other. Both types of excitations have a gap, i.e. for $g_1 < 0$ one has a *gap in the spin excitation spectrum*, whereas the charge excitations remain massless. This can actually investigated in more detail in an exactly solvable case [Luther and Emery, 1974].

---

[*]The $N_\nu$- and $J_\nu$–terms are discussed by Haldane [Haldane, 1981a].

In a half–filled band, umklapp scattering is possible, transferring two particles from $-k_F$ to $k_F$, or vice versa. These processes lead to an extra term

$$\frac{2g_1}{(2\pi\alpha)^2} \int dx \cos(\sqrt{8}\phi_\rho) \tag{3.7}$$

in the Hamiltonian. Similarly to the $g_1$ term, this can lead to a gap, this time in the charge excitation spectrum [Emery et al., 1976, Sólyom, 1979]. The ground state then is insulating. This happens in particular for the Hubbard model with repulsive interaction at exactly half–filling.

## 3.2 Spin–Charge Separation

One of the more spectacular consequences of the Hamiltonian (3.3) is the complete separation of the dynamics of the spin and charge degrees of freedom. For example, in general one has $u_\sigma \neq u_\rho$, i.e. the charge and spin oscillations propagate with different velocities. Only in a noninteracting system one has $u_\sigma = u_\rho = v_F$. To make the meaning of this fact more transparent, let us create an extra particle in the ground state, at $t = 0$ and spatial coordinate $x_0$. The charge and spin densities then are easily found, using $\rho(x) = -(\sqrt{2}/\pi)\partial\phi_\rho/\partial x$ (note that $\rho(x)$ is the deviation of the density from its average value) and $\sigma_z(x) = -(\sqrt{2}/\pi)\partial\phi_\sigma/\partial x$ :

$$\begin{aligned}
\langle 0|\psi_+(x_0)\rho(x)\psi_+^\dagger(x_0)|0\rangle &= \delta(x-x_0) \ , \\
\langle 0|\psi_+(x_0)\sigma_z(x)\psi_+^\dagger(x_0)|0\rangle &= \delta(x-x_0) \ .
\end{aligned} \tag{3.8}$$

Now, consider the time developement of the charge and spin distributions. The time–dependence of the charge and spin density operators is easily obtained from (3.3) (using the fixed point value $g_1 = 0$), and one obtains

$$\begin{aligned}
\langle 0|\psi_+(x_0)\rho(x,t)\psi_+^\dagger(x_0)|0\rangle &= \delta(x-x_0-u_\rho t) \ , \\
\langle 0|\psi_+(x_0)\sigma_z(x,t)\psi_+^\dagger(x_0)|0\rangle &= \delta(x-x_0-u_\sigma t) \ .
\end{aligned} \tag{3.9}$$

Because in general $u_\sigma \neq u_\rho$, after some time charge and spin will be localized at completely different points in space, i.e. *charge and spin have separated completely.* A interpretation of this surprising phenomenon in terms of the Hubbard model will be given in sec.(4). Here a linear energy–momentum relation has been assumed for the electrons, and consequently the shape of the charge and spin distributions is time–independent. If the energy–momentum relation has some curvature (as is necessarily the case in lattice systems) the distributions will widen with time. However this widening is proportional to $\sqrt{t}$, and therefore much smaller than the distance between charge and spin. Thus, the qualitative picture of spin-charge separation is unchanged.

## 3.3 Luttinger Liquid

The simple form of the Hamiltonian (3.3) at the fixed point $g_1^* = 0$ makes the calculation of physical properties rather straightforward. First note that acoustic phonons in one dimension have a linear specific heat. Consequently, the low-temperature specific heat of interacting fermions is $C(T) = \gamma T$, with

$$\gamma/\gamma_0 = \frac{1}{2}(v_F/u_\rho + v_F/u_\sigma) \ . \tag{3.10}$$

Here $\gamma_0$ is the specific heat coefficient of noninteracting electrons of Fermi velocity $v_F$.

The spin susceptibility and the compressibility are equally easy to obtain. Note that in (3.3) the coefficient $u_\sigma/K_\sigma$ determines the energy necessary to create a nonzero spin polarization, and $u_\rho/K_\rho$ fixes the energy needed to change the particle density. Given the fixed point value $K_\sigma^* = 1$, one finds

$$\chi/\chi_0 = v_F/u_\sigma \ , \kappa/\kappa_0 = v_F K_\rho/u_\rho \ , \tag{3.11}$$

where $\chi_0$ and $\kappa_0$ are the susceptibility and compressibility of the noninteracting case. From eqs.(3.10) and (3.11) the Wilson ratio is

$$R_W = \frac{\chi}{\gamma}\frac{\gamma_0}{\chi_0} = \frac{2u_\rho}{u_\rho + u_\sigma} \ . \tag{3.12}$$

The quantity $\Pi_\rho(x)$ is proportional to the current density. Obviously, the Hamiltonian commutes with the total current, and therefore the frequency dependent conductivity is a delta function at $\omega = 0$. Using the Kubo formula, one straightforwardly finds

$$\sigma(\omega) = 2K_\rho u_\rho \delta(\omega) \ , \tag{3.13}$$

i.e. the product $K_\rho u_\rho$ determines the weight of the dc peak in the conductivity.

The above properties, linear specific heat, finite spin susceptibility, and dc conductivity are those of an ordinary Fermi liquid, the coefficients $u_\rho, u_\sigma$, and $K_\rho$ determining renormalizations with respect to noninteracting quantities. We will now consider quantities which show that a one-dimensional interacting fermion system is *not* a Fermi liquid. Consider the single-particle Green's function which can be calculated using the representation (3.1) of fermion operators. One then finds for the momentum distribution function in the vicinity of $k_F$:

$$n_k \approx n_{k_F} - \beta\text{sign}(k - k_F)|k - k_F|^\alpha \ , \tag{3.14}$$

and for the single-particle density of states: $N(\omega) \approx |\omega|^\alpha$, with $\alpha = (K_\rho + 1/K_\rho - 2)/4$, and $\beta$ is a model-dependent constant. Note that for any $K_\rho \neq 1$, i.e. *for any nonvanishing interaction*, the momentum distribution function and the density of states have power-law singularities at the Fermi level, with a vanishing single particle density of states at $E_F$. This behaviour is obviously quite different from a standard Fermi liquid which would have a finite density of states and a step-like singularity in $n_k$. The absence of a step at $k_F$ in the momentum distribution function implies the *absence of a quasiparticle pole* in the one-particle Green's function (this can of course be verified directly). The unusual type of behaviour found here has been called **Luttinger liquid** by Haldane [Haldane, 1981a].

The coefficient $K_\rho$ also determines the long-distance decay of all other correlation functions of the system: Using the representation (3.1) the operators (2.14) become

$$O_{CDW}(x) = \lim_{\alpha \to 0} \frac{e^{2ik_F x}}{\pi \alpha} e^{-i\sqrt{2}\phi_\rho(x)} \cos[\sqrt{2}\phi_\sigma(x)] \ , \tag{3.15}$$

$$O_{SDW_x}(x) = \lim_{\alpha \to 0} \frac{e^{2ik_F x}}{\pi \alpha} e^{-i\sqrt{2}\phi_\rho(x)} \cos[\sqrt{2}\theta_\sigma(x)] \ , \tag{3.16}$$

where $\theta_\nu(x) = \pi \int^x \Pi_\nu(x') dx'$. Similar relations are also found for the other operators in (2.14). It is important to note here that all these operators decompose into a product of one factor depending on the charge variable only by another factor depending only on the spin field. Using the Hamiltonian (3.3) at the fixed point $g_1^* = 0$ one finds for example for the charge and spin correlation functions[†]

$$\langle n(x)n(0) \rangle = K_\rho/(\pi x)^2 + A_1 \cos(2k_F x) x^{-1-K_\rho} \ln^{-3/2}(x)$$
$$+ A_2 \cos(4k_F x) x^{-4K_\rho} + \ldots \ , \tag{3.17}$$

$$\langle \vec{S}(x) \cdot \vec{S}(0) \rangle = 1/(\pi x)^2 + B_1 \cos(2k_F x) x^{-1-K_\rho} \ln^{1/2}(x) + \ldots \ , \tag{3.18}$$

with model dependent constants $A_i, B_i$. The ellipses in (3.17) and (3.18) indicate higher harmonics of $\cos(2k_F x)$ which are present but decay faster than the terms shown here. Similarly, correlation functions for singlet (SS) and triplet (TS) superconducting pairing are

$$\langle O_{SS}^\dagger(x) O_{SS}(0) \rangle = C x^{-1-1/K_\rho} \ln^{-3/2}(x) + \ldots \ ,$$
$$\langle O_{TS_\alpha}^\dagger(x) O_{TS_\alpha}(0) \rangle = D x^{-1-1/K_\rho} \ln^{1/2}(x) + \ldots \ . \tag{3.19}$$

The logarithmic corrections in thses functions have been found rather recently [Giamarchi and Schulz, 1989]. The corresponding susceptibilities (i.e. the Fourier transforms of the above correlation functions) behave at low temperatures as

$$\chi_{CDW}(T) \approx T^{K_\rho - 1} |\ln(T)|^{-3/2} \ , \chi_{SDW}(T) \approx T^{K_\rho - 1} |\ln(T)|^{1/2} \ , \tag{3.20}$$

$$\chi_{SS}(T) \approx T^{1/K_\rho - 1} |\ln(T)|^{-3/2} \ , \chi_{TS}(T) \approx T^{1/K_\rho - 1} |\ln(T)|^{1/2} \ , \tag{3.21}$$

i.e. for $K_\rho < 1$ (spin or charge) density fluctuations at $2k_F$ are enhanced and diverge at low temperatures, whereas for $K_\rho > 1$ pairing fluctuations dominate. The remarkable fact in all the above results uis that there is only *one coefficient*, $K_\rho$, which determines all the asymptotic power laws.

If in a half–filled band umklapp scattering leads to an insulating state with a gap in the charge excitation spectrum, the above results are valid with $K_\rho = 0$ (then $x^{-1/0}$ indicates exponential decay).

---

[†]The time- and temperature dependence is also easily obtained, see [Emery, 1979].

# 4  The Hubbard model in one dimension

## 4.1  The Hamiltonian and its symmetries

The Hubbard model is the protypical model used for the description of correlated fermions in a large variety of circumstances, ranging from high–$T_c$ superconductors to heavy fermion compounds and organic conductors. In spite of its apparent simplicity, there is still no general solution, or even a consensus on its fundamental properties. Notable exceptions are the cases of one and infinite dimensions [Vollhardt, 1992, Georges and Kotliar, 1992]. In particular, in one dimension an exact solution is available. This solution gives exact energies of the ground state and all the excited states in terms of the solution of a system of coupled nonlinear equations. On the other hand, the corresponding wavefunctions have a form so complicated that the explicit calculation of matrix elements, correlation functions and other physical quantities has remained impossible so far. In the following sections I shall describe the features of the one–dimensional Hubbard model which make it exactly solvable, and subsequently describe in detail the energy spectrum. Finally, I will show how the knowledge of the energy spectrum can be combined with the results of the preceding two chapters to obtain a rather detailed picture of the low–energy properties, in particular of correlation functions.

The Hamiltonian in one dimension has the well–known form

$$H = -t \sum_{i,s}(c^\dagger_{i,s}c_{i+1,s} + c^\dagger_{i+1,s}c_{i,s}) + U\sum_i n_\uparrow n_\downarrow \; , \qquad (4.1)$$

where $c_{i,s}$ is the fermion annihilation operator on site $i$ with spin $s$, $n_{i,s}$ is the corresponding number operator, and the sum is over the $L$ sites of a one–dimensional chain with periodic boundary conditions.

The model has *two* global SU(2) symmetries [Pernici, 1990, Zhang, 1990, Schulz, 1990a]: the first is the well–known spin rotation invariance, with generators

$$\zeta = \sum_{i=1}^L c^\dagger_{i,\uparrow}c_{i,\downarrow}, \quad \zeta^\dagger = (\zeta)^\dagger, \quad \zeta_z = \frac{1}{2}\sum_{i=1}^L (n_{i,\downarrow} - n_{i,\uparrow}) \; . \qquad (4.2)$$

The second type of symmetry is particular to the Hubbard model and relates sectors of different particle numbers. Its generators are

$$\eta = \sum_{i=1}^L (-1)^i c_{i,\uparrow} c_{i,\downarrow}, \quad \eta^\dagger = (\eta)^\dagger, \quad \eta_z = \frac{1}{2}\sum_{i=1}^L (n_{i,\downarrow} + n_{i,\uparrow}) - \frac{L}{2} \; . \qquad (4.3)$$

The total symmetry thus is $SU(2) \times SU(2) \simeq SO(4)$. One should notice that more complicated interactions, e.g. involving further neighbours, will conserve the spin rotation invariance but in general not the "charge" SU(2) invariance (4.3). Rather, this second symmetry will become the standard global $U(1)$ invariance associated with particle number conservation. It is nevertheless possible to construct particular types of further–neightbour interactions which do conserve the full $SU(2) \times SU(2)$ invariance.

## 4.2 The exact solution: a brief introduction

I here closely follow a presentation by N. Andrei [Andrei, 1988]. To obtain the wavefunctions of the Hubbard model, one writes a general $N$–particle state as

$$|F\rangle = \sum_{x_1}\ldots\sum_{x_N}\sum_{s_1}\ldots\sum_{s_N} F_{s_1,\ldots,s_N}(x_1,\ldots,x_N) \prod_{i=1}^{N} c^{\dagger}_{x_i,s_i}|vac\rangle \ . \tag{4.4}$$

The simplest case is the two–particle problem. Then in the two parts of configuration space $x_2 > x_1$ (region I) and $x_1 > x_2$ (region II) the wavefunction is just a product of two plane waves, and the only nontrivial effects occur for $x_1 = x_2$. The full wavefunction $F$ then is

$$F_{s_1,s_2} = e^{i(k_1 x_1 + k_2 x_2)}[\theta(x_2 - x_1)\xi^I_{s_1,s_2} + \theta(x_1 - x_2)\xi^{II}_{s_1,s_2}] \ , \tag{4.5}$$

where the factors $\xi$ depend only on the spin quantum numbers of the two particles. In order to satisfy the Schrödinger equation, these coefficients have to obey the equation

$$\xi^{II}_{s_1,s_2} = \sum_{t_1,t_2} S_{s_1 s_2, t_1 t_2} \xi^I_{t_1 t_2} \ . \tag{4.6}$$

The $S$–matrix has the form

$$S_{\alpha\beta,\gamma\delta} = \frac{t(\sin k_1 - \sin k_2) + iU P_{\alpha\beta,\gamma\delta}}{t(\sin k_1 - \sin k_2) + iU} \ , \tag{4.7}$$

where $P_{\alpha\beta,\gamma\delta}$ is the operator permuting two spins. This operator acts on the part of the two–particle Hilbert space.

For $N$ particles, there are $N!$ different regions in configuration space, each correspond to a permutation $Q$ of the sequence $(1,2,3,\ldots,N)$, so that $x_{Q1} < x_{Q2} < \ldots < x_{QN}$. The wavefunction takes the form

$$F_{s_1,\ldots,s_N}(x_1,\ldots,x_N) = e^{i\sum_j k_j x_j} \sum_{Q \in S_N} \theta_Q[x_\alpha] \xi_{s_1,\ldots,s_N}(Q) \ , \tag{4.8}$$

where $S_N$ is the group of permutations of $N$ objects, and $\theta_Q[x_\alpha] = 1$ if $x_{Q1} < x_{Q2} < \ldots < x_{QN}$ and zero otherwise.. The Schrödinger equation imposes the condition

$$\xi(Q') = S^{ij}\xi(Q) \tag{4.9}$$

when $Q$ and $Q'$ are two permutations which only differ by the exchange $x_i \leftrightarrow x_j$.

However, for more than two particles, there is more than one way to go from one part of configuration space to another. Consider three particles. Denoting by $(ijk)$ the part of configuration space with $x_i < x_j < x_k$, there are obviously the possibilities

$$(123) \to (213) \to (231) \to (321)$$

and

$$(123) \to (132) \to (312) \to (321)$$

In order for the two paths to lead to the same amplitude in (321), one has to have

$$S^{23}S^{13}S^{12} = S^{12}S^{13}S^{23} \;. \tag{4.10}$$

These are the famous *Yang–Baxter equations* [Yang, 1967] which have to be satisfied for a system to be solvable by Bethe Ansatz. One can verify that these equations are indeed satisfied by the S–matrix of the Hubbard model, eq. (4.7).

One then imposes periodic boundary conditions. This means that the relation

$$F_{s_1,\ldots,s_N}(x_1,\ldots,x_j=0,\ldots) = F_{s_1,\ldots,s_N}(x_1,\ldots,x_j=L,\ldots) \tag{4.11}$$

should be valid for every $j$. When $x_j$ varies from 0 to $L$ two things happen: (i) the wavefunction is multiplied by a factor $e^{ik_j L}$, and (ii) particle $j$ is permuted with all the other particles. To satisfy periodic boundary conditions, one therefore needs to satisfy the relation

$$Z_j \xi_{s_1,\ldots,s_N}(I) = e^{ik_j L} \xi_{s_1,\ldots,s_N}(I) \;, \tag{4.12}$$

where $Z_j$, acting on the spin Hilbert space of $N$ particles, is an operator representing the phase shifts as particle $j$ crosses all the other particles:

$$Z_j = S^{j,j-1} \ldots S^{j,1} S^{j,N} \ldots S^{j,j+1} \tag{4.13}$$

Thus, the spin wavefunctions $\xi$ are the eigenfunctions of the operator $Z_j$, and the allowed values of the momenta $k$ are related to the eigenvalues of $Z_j$. The determination of the eigenfunctions of $Z_j$ is related to the so–called 6–vertex model of statistical mechanics. A detailed description of the method used can be found in more specialized articles [Baxter, 1982, Thacker, 1982].

Using those results, the allowed values of $k_j$ are obtained from the solution of the coupled set of nonlinear equations

$$e^{ik_j L} = \prod_{\alpha=1}^{M} e\left(\frac{4(\sin k_j - \lambda_\alpha)}{U}\right) \tag{4.14}$$

$$\prod_{j=1}^{N} e\left(\frac{4(\lambda_\alpha - \sin k_j)}{U}\right) = -\prod_{\beta=1}^{M} e\left(\frac{2(\lambda_\alpha - \lambda_\beta)}{U}\right) \;, \tag{4.15}$$

Here $N$ is the total number of electrons, $M$ is the number of down–spin electrons ($M \leq N/2$), and $e(x) = (x+i)/(x-i)$. The $\lambda_\alpha$ are parameters characterizing the spin dynamics. We note that in general, both the $k_j$'s and the $\lambda$'s are allowed to be complex. The energy and momentum of a state are

$$E = -2t\sum_{j=1}^{N} \cos k_j \;, \qquad P = \sum_{j=1}^{N} k_j \;. \tag{4.16}$$

### 4.2.1 Solutions of the Bethe ansatz equations

The determination of all the solutions of eqs. (4.14, 4.15) is not easy. It has recently been shown (under certain assumptions) that these equations do indeed give all the "lowest weight" (with respect to $SU(2) \times SU(2)$) eigenstates of the Hubbard model, i.e. all states satisfying $\eta|\psi\rangle = \zeta|\psi\rangle = 0$. The complete set of eigenstates then is obtained acting repeatedly with $\eta^\dagger$ or $\zeta^\dagger$ on $|\psi\rangle$ [Essler et al., 1991]. Here I will limit myself to the ground state and to the low–lying elementary excitations. These questions have been investigated in some detail [Shiba, 1972, Coll, 1974, Woynarovich, 1983], however, the finite chain data presented below seem to be quite useful in understanding the nature of the excitations, and I therefore discuss them in detail.

If both the $k$'s and the $\lambda$'s are all real, only the phases in (4.14, 4.15) have to be determined. Taking the logarithm of these equations, one finds

$$Lk_j = 2\pi I_j + 2 \sum_{\alpha=1}^{M} \arctan[4(\lambda_\alpha - \sin k_j)/U] \tag{4.17}$$

$$2 \sum_{j=1}^{N} \arctan[4(\lambda_\alpha - \sin k_j)/U] = 2\pi J_\alpha + 2 \sum_{\beta=1}^{M} \arctan[2(\lambda_\alpha - \lambda_\beta)/U] \; . \tag{4.18}$$

The quantum numbers $\{I_j\}$ are all distinct from each other and are integers if $M$ is even and half–odd integers (HOI, i.e. of the form 1/2, 3/2, ...) if $M$ is odd, and are only defined modulo $L$. Similarly, the set $\{J_\alpha\}$ are all distinct and are integers if $N - M$ is odd and HOI if $N - M$ is even. Moreover, there is the restriction

$$|J_\alpha| < (N - M + 1)/2 \; . \tag{4.19}$$

Summing (4.17) over $j$ and (4.18) over $\alpha$, the total momentum is found as

$$P = \frac{2\pi}{L} \left( \sum_{j=1}^{N} I_j + \sum_{\alpha=1}^{M} J_\alpha \right) \; . \tag{4.20}$$

**Ground state.** The ground state is nondegenerate only if $N$ is of the form $4\nu + 2$ ($\nu$ integer): obviously, if $N$ is odd, the ground state has (at least) spin 1/2. Further, if $N$ is an integer multiple of 4 the noninteracting ground state has a sixfold degeneracy, and in the interacting case the ground state turns out to be a spin triplet. In the following I shall restrict myself to the case of $N = N_0 = 4\nu + 2$, i.e. the ground state is nondegenerate (in the following, $N_0$ will denote the particle number in the ground state). The ground state then is a singlet, with $M = N_0/2$, i.e. $M$ is odd. The allowed values of the $J$'s range from $-(N_0/2 - 1)/2$ to $(N_0/2 - 1)/2$. There are exactly $N_0/2$ such integers, i.e. all the $J$'s are fixed. The $I$'s are consecutive between $-(N_0 - 1)/2$ and $(N_0 - 1)/2$, i.e.

$$\{I_j\} = \{-(N_0 - 1)/2, \ldots, (N_0 - 1)/2\} \; , \tag{4.21}$$

$$\{J_\alpha\} = \{-(N_0/2 - 1)/2, \ldots, (N_0/2 - 1)/2\} \; . \tag{4.22}$$

In the thermodynamic limit $L \to \infty$ the distance between consecutive $k$'s or $\lambda$'s decrease like $1/L$, and one can then find linear integral equations for the density of $k$'s and $\lambda$'s on the real axis. Numerical results for the ground state energy as a function of particle density and $U$ have been given by Shiba [Shiba, 1972].

**"$4k_F$" singlet states.** Excited states are obtained by varying the quantum numbers. The first possibility, giving rise to excited singlet states, is obtained by removing one of the $I$'s from the ground state sequence (4.21) and adding a "new" $I$:

$$\begin{aligned}
\{I_j\} &= \{-(N_0-1)/2, \ldots, -(N_0-1)/2 + i_0 - 1, \\
&\qquad -(N_0-1)/2 + i_0 + 1, \ldots, (N_0-1)/2, I_0\} \;, \\
\{J_\alpha\} &= \{-(N_0/2-1)/2, \ldots, (N_0/2-1)/2\} \;,
\end{aligned} \qquad (4.23)$$

where $|I_0| > (N_0-1)/2$. This is a two-parameter family of excited states, called (somewhat misleadingly) "particle–hole excitation" by Coll. To understand the excited states, we shall in the following consider systems of finite size, rather than taking the thermodynamic limit directly. It should however be quite evident how spectra like those of fig.5 develop into a true continuum in the thermodynamic limit (compare in particular figs.6 and 7). In fig.5 we show numerical results for the energy–momentum spectrum of the states (4.23) for a chain of 40 sites. The same states with $k \to -k$ are obtained using negative $I_0$. One notices a sharp minimum in the exctation energy at $k/\pi = 1.1 = 4k_F$ (this is why I call these excitations "$4k_F$" singlts). In the thermodynamic limit, the gap at $4k_F$ vanishes. These excitations are at the origin of the power–law behaviour in the density-density correlation function (3.17) around $4k_F$. Moreover, in the bosonization formalism, the $4k_F$ density correlations are entirely determined by the charge ($\phi_\rho$) modes. Consequently, we identify the charge velocity $u_\rho$ as the slope of the excitation spectrum of fig.5 at $k = 0$. Finally, we notice that the overall structure of the spectrum doesn't change much between weak ($U/t = 1$) and strong ($U/t = 16$) correlation cases.

**"$2k_F$" triplet and singlet states.** Excitations of the $J$'s with all $\lambda$'s and $k$'s real are only possible if $M < N/2$. The simplest excitations of this type are obtained considering $M = N/2 - 1$ which has total spin $S = 1$ (triplet). Now the restriction (4.19) allows $N/2 + 1$ different $J$'s, i.e. we have two free parameters (the "holes" in the $J$–sequence), leading to sequences of quantum numbers of the form

$$\begin{aligned}
\{I_j\} &= \{-N_0/2 + 1, \ldots, N_0/2\} \;, \\
J_1 &= -N_0/4 + \delta_{\alpha_1,1} \;, \\
J_\alpha &= J_{\alpha-1} + 1 + \delta_{\alpha,\alpha_1} + \delta_{\alpha,\alpha_2} \quad (\alpha = 2, \ldots, M) \;,
\end{aligned} \qquad (4.24)$$

where $1 \le \alpha_1 < \alpha_2 \le M$, and $\delta_{\alpha,\beta}$ is the usual Kronecker symbol. Numerical results for this type of excitations are shown in fig.6. Corresponding states with negative $k$ are obtained shifting the $\{I_j\}$ in (4.24) by one unit to the left. There now is a sharp minimum at $k/\pi = 0.55 = 2k_F$, the gap again vanishing in the thermodynamic limit. In the long–wavelength limit, these states are the only spin–carrying excitations at constant particle number, so that the slope of the spectrum at $k = 0$ is equal to the spin velocity of the bosonized model. The low–energy excitations around $2k_F$ are responsible for the spin contribution to the $2k_F$ spin–spin correlations. As in fig.5, the structure of the excitations doesn't change much between weak and strong correlations, however, the energy scale does. In fact, the lowering of the energy scale in going to strong correlations corresponds to the lowering of the exchange energy ($\approx 4t^2/U$) in the strong correlation limit.

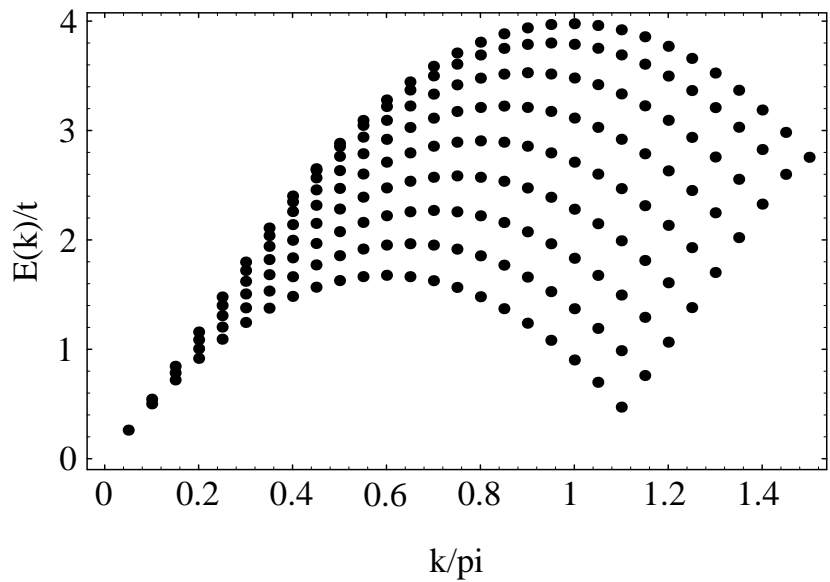

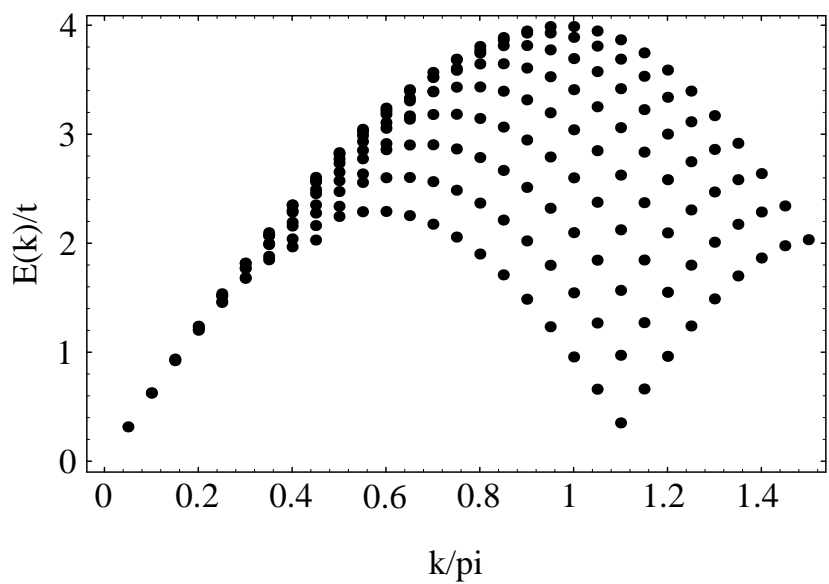

Figure 5: "$4k_F$" singlet excitation spectrum for a Hubbard chain of 40 sites with 22 electrons. The lowest "arc" from $k/\pi = 0.05$ to $k/\pi = 1.1$ is obtained by varying $i_0$ at fixed $I_0 = (N_0 + 1)/2$ (cf. eq. (4.23)), the higher arches correspond to increasing $I_0$ up to $(L-1)/2$.

The results of fig.6 show apparent gaps at $2k_F$ and $4k_F$. These are clearly finite size effects, as can be seen comparing with fig.7 which contains results for the triplet excitations for the same particle density and interaction strength as in fig.6, but for a chain four times longer. The gaps are now much smaller and obey approximately the $1/L$ scaling expected. One also easily recognizes how the spectra develop into a continuum in the thermodynamic limit.

Together with the triplet excitations (4.24) there are also singlet states ($M = N_0/2$), which are obtained by having one pair of complex conjugate $\lambda$'s among the solutions to the original equations (4.14, 4.15). The energies of these states are shown by triangles in fig.5. It is remarkable that these states are nearly degenerate with the triplet states and in fact become exactly degenerate in the thermodynamic limit.

The existence of singlets and triplets with the same energy shows that these states are in fact the combination of two *noninteracting spin–1/2 objects*, commonly called *spinons*. Of course, because of total spin conservation, these objects can only be excited in pairs as long as one keeps the total number of particles fixed.

**Added particle.** Adding one particle to the $4\nu + 2$ ground state and leaving $M$ unchanged, the $I$'s and the $J$'s are HOI. There are now $M + 1$ allowed values for the $M$ distinct $J$'s, and the low-energy states then are parametrized by

$$\begin{aligned}
\{I_j\} &= \{-(N_0 - 1)/2, \ldots, (N_0 - 1)/2, I_0\} \ , \\
J_1 &= -M/2 + \delta_{\alpha_1, 1} \ , \\
J_\alpha &= J_{\alpha-1} + 1 + \delta_{\alpha, \alpha_1} \quad (\alpha = 2, \ldots, M) \ ,
\end{aligned} \quad (4.25)$$

where $|I_0| > (N_0 - 1)/2$, and $1 \leq \alpha_1 \leq M$. Corresponding spectra for different bandfillings are shown in fig.8 for $I_0 > 0$. The symmetric spectra with negative $k$ are again obtained using $I_0 < 0$. The state of minimal excitation energy has momentum $k_F$, as in the noninteracting case. The shallow arches in fig.8 correspond to varying $\alpha_1$, and are in fact of the same shape as the lowest branch (from $k = 0.05\pi$ to $k = 0.55\pi$) in fig.6, i.e. they correspond to *single–spinon* excitations. Close to $k_F$ the energy of theses states varies as $u_\sigma(k - k_F)$. On the other hand, varying $I_0$, one goes from one arc to the next, and the corresponding excitation energy (the upper limit of the quasi–continuum) varies as $u_\rho(k - k_F)$. One also sees that going from one arc to the next in fig.8 (i.e. increasing $I_0$) the shape of an individual arc is basically unchanged. Varying $I_0$ corresponds to a variation of the momentum of the added particle, and the figure thus shows that the total energy of a state is just the sum of the spinon energy and the "charge" energy associated with the added particle. One thus sees that *charge and spin degrees of freedom do not interact*. This is certainly in agreement with the predictions of the bosonization formalism, however, the fact that spin and charge separate even in highly excited states is special to the Hubbard model (a priori, the bosonized theory can only be expected to be an effective low–energy theory for the Hubbard and other lattice models).

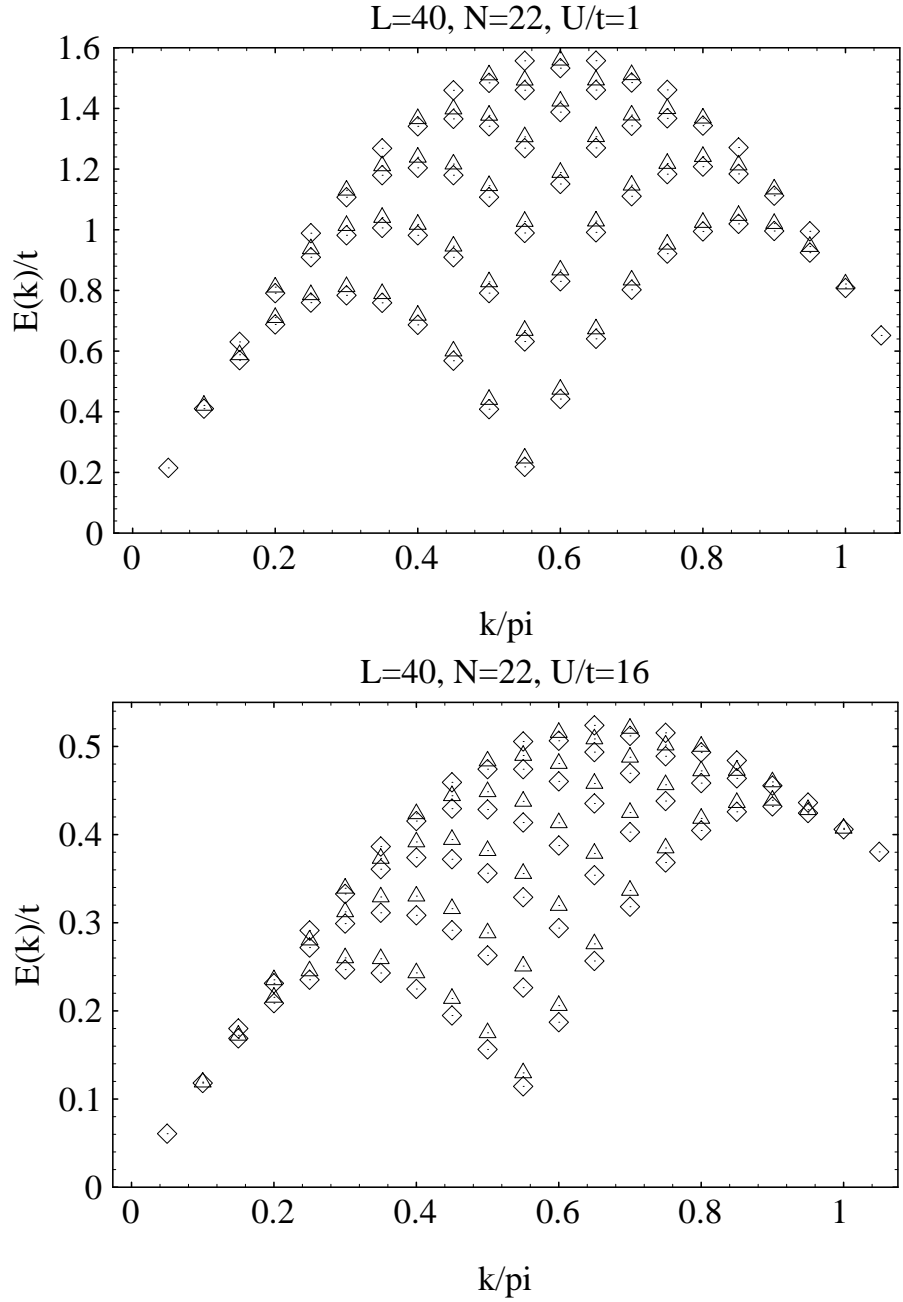

Figure 6: "$2k_F$" spin singlet ($\triangle$) and triplet ($\diamond$) excitation spectrum for Hubbard chains of 40 sites with 22 electrons for different interaction strengths.

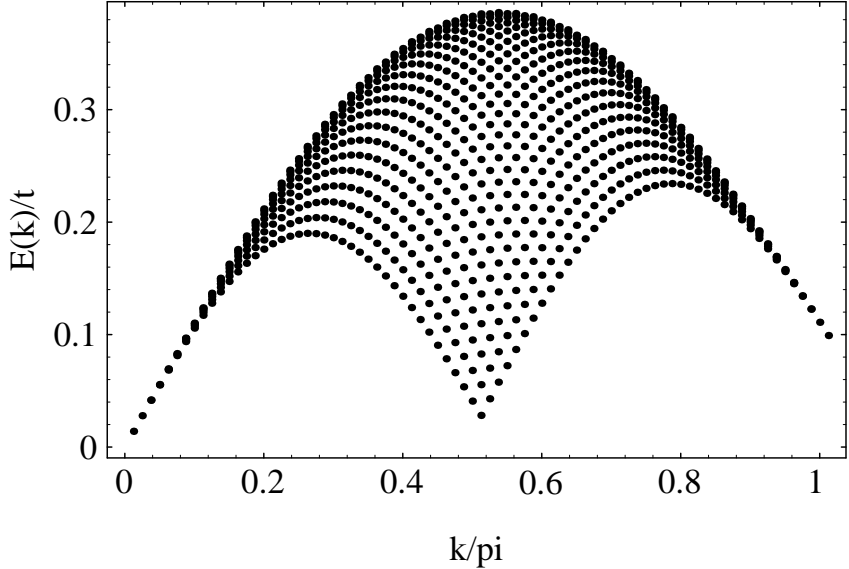

Figure 7: "$2k_F$" triplet excitation spectrum for a Hubbard chain of 160 sites with 82 electrons.

Another notable fact in fig.8 is the number of available states as $I_0$ is varied: for $N_0 = 14, 22$, and 30 there are respectively $13, 9$, and 5 spinon arches. This means that without exciting the spins there are $26, 18$, and 10 states for the extra particle available (counting states at negative $k$), i.e. the $I_0$ branch stops at $k = \pi - k_F$, rather than at $k = \pi$ as in a noninteracting system. The explanation of this fact is rather straightforward for large $U$ when double occupancy of sites is forbidden: in a system with $L$ sites and $N_0$ electrons, there are only $L - N_0$ sites available at low energies. These states then form the "band" in fig.8. There are of course states involving doubly occupied sites, however these are separated from the continuum of fig.8 by a gap (these states are solutions of (4.14, 4.15) with complex $k$'s [Woynarovich, 1982]). This separation of states occurs for any, even very small $U$.

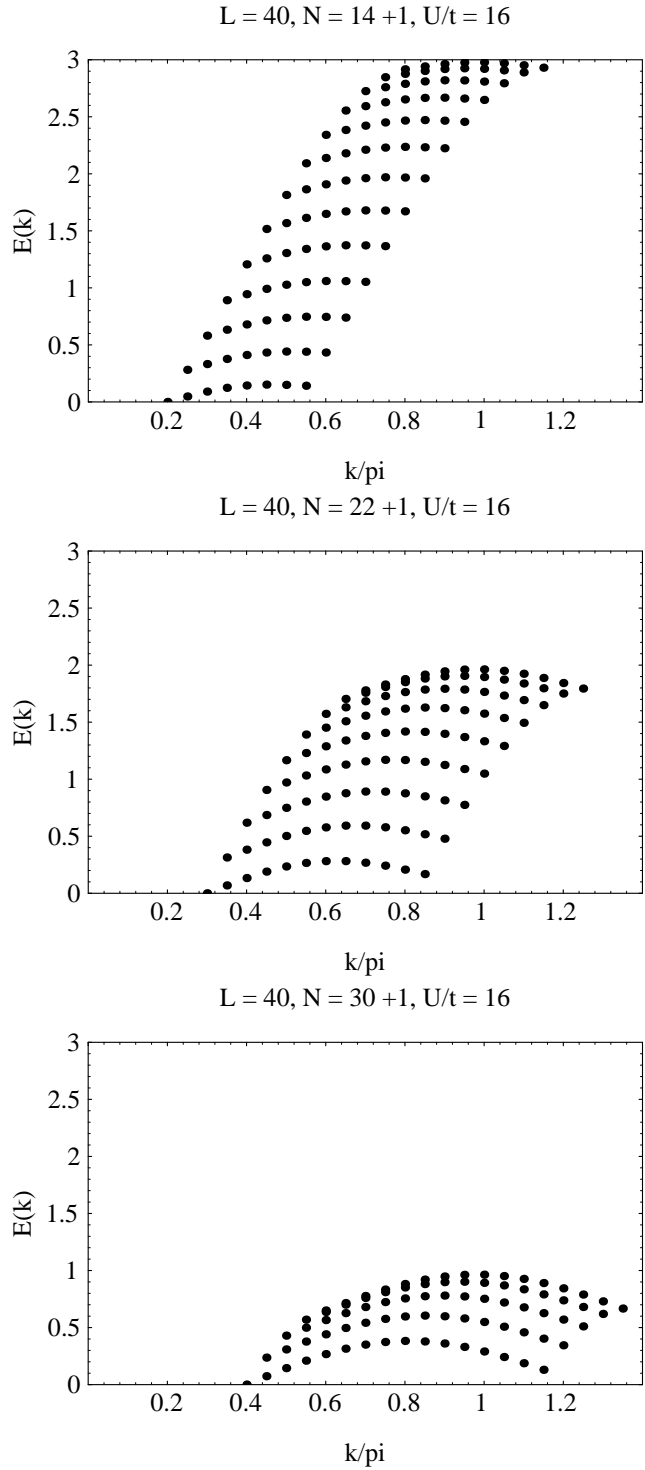

Figure 8: *Excitation spectra for one particle added into a Hubbard chain of 40 sites with 14, 22, and 30 electrons. The shallow arches correspond to varying $\alpha_1$ (cf. eq. (4.25)) at constant $I_0$, and $I_0$ increases from one arc to the next. Zero energy corresponds to the $N_0 + 1$ particle ground state.*

**Added hole.** Finally, let us consider states with one hole in the $4\nu + 2$ groundstate ($N = N_0 - 1, M = (N-1)/2$). Then both the $I$'s and the $J$'s are integers. The energy is minimized chosing consecutive $I$'s between $-(N-1)/2$ and $(N-1)/2$, but there are $M + 1$ possibilities for the $M$ $J$'s. States corresponding to the sequence

$$\begin{align}
\{I_j\} &= \{-(N-1)/2, \ldots, (N-1)/2\} , \\
J_1 &= -M/2 + \delta_{\alpha_1, 1} , \\
J_\alpha &= J_{\alpha-1} + 1 + \delta_{\alpha_1, \alpha} \quad (\alpha = 2, \ldots, M)
\end{align} \tag{4.26}$$

are shown as the lowest arc between $k = -0.25\pi$ and $k = 0.25\pi$. This is a one–spinon branch, this state having necessarily $S = 1/2$, with velocity $u_\sigma$. One can of course also create a hole in the sequence of $I$'s. The energy spectrum for the quantum numbers

$$\begin{align}
I_1 &= -(N+1)/2 + \delta_{j_1, 1} \\
I_j &= I_{j-1} + 1 + \delta_{j_1, j} \quad (j = 2, \ldots, N) \tag{4.27} \\
J_1 &= -M/2 + \delta_{\alpha_1, 1} , \\
J_\alpha &= J_{\alpha-1} + 1 + \delta_{\alpha_1, \alpha} \quad (\alpha = 2, \ldots, M) \tag{4.28}
\end{align}$$

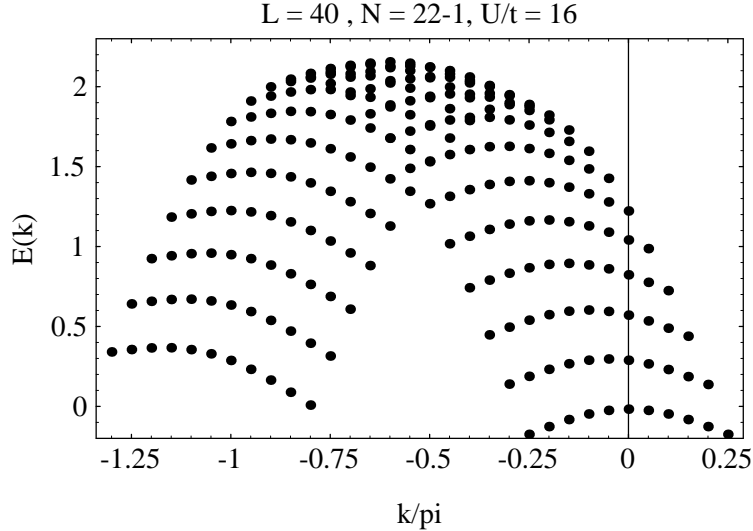

Figure 9: *Excitation spectra for an added hole in a Hubbard chain of 40 sites with 22 electrons. The shallow arches correspond to varying $\alpha_1$ (cf. eq. (4.28) at constant $j_1$, and $j_1$ increases from one arc to the next.*

(where the free parameters obey $1 \leq j_1 \leq N$, $1 \leq \alpha_1 \leq M$) is also shown in fig.9. Similarly to fig.8, one notices that varying the "charge" quantum number $j_1$ one creates a branch with velocity $u_\rho$.

### 4.2.2 Limiting cases

**Strong correlation.** For $U \to \infty$ the ratio $\sin k_j/U$ in (4.14,4.15) clearly vanishes, howver there is no such restriction on the $\lambda$'s. Introducing the scaled variables $\Lambda_\alpha = 2\lambda_\alpha/U$, the "spin equation" (4.15) becomes

$$\left(\frac{2\Lambda_\alpha + i}{2\Lambda_\alpha - i}\right)^N = -\prod_{\beta=1}^{M} \frac{\Lambda_\alpha - \Lambda_\beta + i}{\Lambda_\alpha - \Lambda_\beta - i} \ , \tag{4.29}$$

which are the well–known Bethe ansatz equations for the spin–1/2 Heisenberg chain [Faddeev and Takhtajan, 1981], i.e. the spin wavefunction of $N$ particles is just that of an $N$–site (not $L$–site!) Heisenberg chain, even when there is less than one particle per site. The "charge equation" (4.14) becomes

$$e^{ik_j L} = e^{iP_H} \ , \quad P_H = \sum_{\alpha=1}^{M}[2\arctan(2\Lambda_\alpha) - \pi] \ , \tag{4.30}$$

where $P_H$ is the momentum of the spin excitations of the $N$–site Heisenberg chain. Eq. (4.30) shows that in the strong correlation limit the allowed $k$'s are quantized in units of $2\pi/L$, as for spinless noninteracting fermions (each $k$ is occupied *once*!). This is of course a consequence of strong correlations: double occupancy is forbidden, i.e. the strong local repulsion plays the role of a Pauli principle acting even for electrons of different spin (this is only true in one dimension). The boundary conditions determinig the allowed $k$'s are determined by the spin momentum $P_H$. In particular, for $4\nu + 2$ particles, the ground state momentum of the spins is an odd multiple of $\pi$, so that the allowed $k$'s are HOI multiples of $2\pi/L$, and consequently the ground state is uniquely determined.

In this limit the total wavefunction separates into a product of a $N$–particle Slater determinant for spinless fermions, describing the charge motion, with the spin wavefunction of the Heisenberg chain. This separation has been used to calculate (numerically) a number of correlation function of the one–dimensional Hubbard model in the strong correlation limit [Ogata and Shiba, 1990]

**Weak correlation.** For $U \to 0$ the arctan functions in (4.17, 4.18) become sign functions, e.g.

$$2\arctan[4(\lambda - \sin k)/U] \to \pi \,\text{sign}(\lambda - \sin k) \ . \tag{4.31}$$

Eq.(4.18) then becomes

$$\begin{aligned}
\sum_{j=1}^{N} \text{sign}(\lambda_\alpha - \sin k_j) &= 2J_\alpha + \sum_{\beta=1}^{M} \text{sign}(\lambda_\alpha - \lambda_\beta) \\
&= 2J_\alpha + 2\alpha - M - 1 \ ,
\end{aligned} \tag{4.32}$$

where in the second line we have assumed (without restriction of generality) that $\{J_\alpha\}$ and therefore $\{\lambda_\alpha\}$ are increasing sequences. Because of the restriction (4.19), the minimum value of the r.h.s. of (4.32) is $-N + 2$, i.e. the smallest $\sin k_j$ is smaller than the smallest $\lambda_\alpha$. From (4.32) one immediately finds

$$\sum_{j=1}^{N}[\text{sign}(\lambda_{\alpha+1} - \sin k_j) - \text{sign}(\lambda_\alpha - \sin k_j)] = 2(J_{\alpha+1} - J_\alpha) + 2 \ , \qquad (4.33)$$

i.e. if $J_{\alpha+1} - J_\alpha = n$ there are exactly $n+1$ solutions $k_j$ satisfying $\lambda_\alpha < \sin k_j < \lambda_{\alpha+1}$. Proceeding in the same way with (4.17), one obtains

$$L(k_{j+1} - k_j) = 2\pi(I_{j+1} - I_j) + \pi \sum_{\alpha=1}^{M}[\text{sign}(\lambda_\alpha - \sin k_{j+1}) - \text{sign}(\lambda_\alpha - \sin k_j)] \ , (4.34)$$

i.e. if there is one $\lambda_\alpha$ satisfying $\sin k_j < \lambda_\alpha < \sin k_{j+1}$ one has

$$k_{j+1} - k_j = 2\pi(I_{j+1} - I_j - 1)/L \ ,$$

whereas if there is no such $\lambda_\alpha$

$$k_{j+1} - k_j = 2\pi(I_{j+1} - I_j)/L \ .$$

From these considerations, the structure of the solution $\{k_j\}, \{\lambda_\alpha\}$ corresponding to the different states above can be obtained in the limit $U \to 0$. For the ground state, one has a sequence $\{\lambda_\alpha\}$, and associated with each $\lambda_\alpha$ there is a pair $(k_{2\alpha-1}, k_{2\alpha})$, so that

$$(\sin k_{2\alpha-1}, \sin k_{2\alpha}) \to \lambda_\alpha \mp \varepsilon \ , \qquad (4.35)$$

and $\varepsilon \to 0$ as $U \to 0$ (see fig.10).

The $U \to 0$ limit of excited states can be discussed similarly. For the $4k_F$ singlets above there are *two* adjacent $k$'s missing in the ground state configuration, one of which is put into the first free state above $k_F$, whereas the other goes into a state of higher momentum, depending on $I_0$ in (4.23). Obviously, for $U = 0$ these states do not contribute to the density correlation function, however, for any nonzero $U$ these two particle/two hole states do couple to the particle–hole excitations, and therefore these states give rise to the $4k_F$ power law in the density correlation function (3.17).

To create the $2k_F$ triplets (4.25) one takes two particles at distinct $k$'s out of the ground state distribution, and puts both into the first available state just above $k_F$. The two unpaired electrons then form a spin triplet, or a singlet if one creates a $2k_F$ singlet. The $U \to 0$ limit of other excited states can be found in the same way. It is important to note here that the simple excited state of the Bethe ansatz discussed above correspond in fact to relatively complicated two particle/two hole states of the noninteracting system.

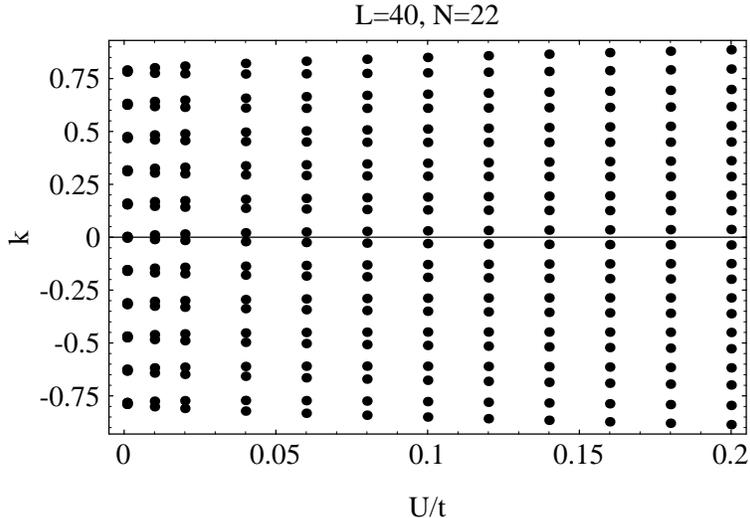

Figure 10: *Dependence of the k's in the groundstate on U. There is one λ between each pair that converges to a doubly occupied state as $U \to 0$.*

## 4.3 Low energy properties of the Hubbard model

### 4.3.1 Luttinger liquid parameters

In a weakly interacting system the coefficients $K_\rho$ and $u_\nu$ can be determined perturbatively. For example, for the Hubbard model one finds

$$K_\rho = 1 - U/(\pi v_F) + \ldots , \qquad (4.36)$$

where $v_F = 2t \sin(\pi n/2)$ is the Fermi velocity for $n$ particles per site. For larger $U$ higher operators appear in the continuum Hamiltonian (3.3), e.g. higher derivatives of the fields or cosines of multiples of $\sqrt{8}\phi_\sigma$. These operators are irrelevant, i.e. they renormalize to zero and do not qualitatively change the long-distance properties, but they do lead to nontrivial corrections to the coefficients $u_\nu, K_\rho$. In principle these corrections can be treated order by order in perturbation theory. However, this approach is obviously unpractical for large $U$, and moreover it is likely that perturbation theory is not convergent. To obtain the physical properties for arbitrary $U$ a different approach is therefore necessary.

I note two points: (i) in the small-$U$ perturbative regime, interactions renormalize to the weak-coupling fixed point $g_1^* = 0, K_\sigma^* = 1$; (ii) the exact solution [Lieb and Wu, 1968] does not show any singular behaviour at nonzero $U$, i.e. large $U$ and small $U$ are the same phase of the model, so that the long-range behaviour even of the large $U$ case is determined by the fixed point $g_1^* = 0$. Thus, the low energy properties of the model are still determined by the three parameters $u_{\rho,\sigma}$ and $K_\rho$.

The velocities $u_{\rho,\sigma}$ can be obtained from the long wavelength limit of the "$4k_F$" and "$2k_F$" excitations discussed above. In the thermodynamic limit the corresponding excitation energies are easily found from the numerical solution of a linear integral equation [Coll, 1974]. Results are shown in fig.11 for various values of $U/t$. Note that for $U = 0$ one has $u_\rho = u_\sigma = 2t\sin(\pi n/2)$, whereas for $U \to \infty$ $u_\rho = 2t\sin(\pi n)$, $u_\sigma = (2\pi t^2/U)(1 - \sin(2\pi n)/(2\pi n))$. In the noninteracting case $u_\sigma \propto n$ for small $n$, but for *any* positive $U$ $u_\sigma \propto n^2$. The Wilson ratio, eq.(3.12) obtained from the velocities is shown in fig.11. For $U = 0$ one has $R_W = 1$, whereas for $U \to \infty$ $R_W = 2$ for $n \neq 1$.

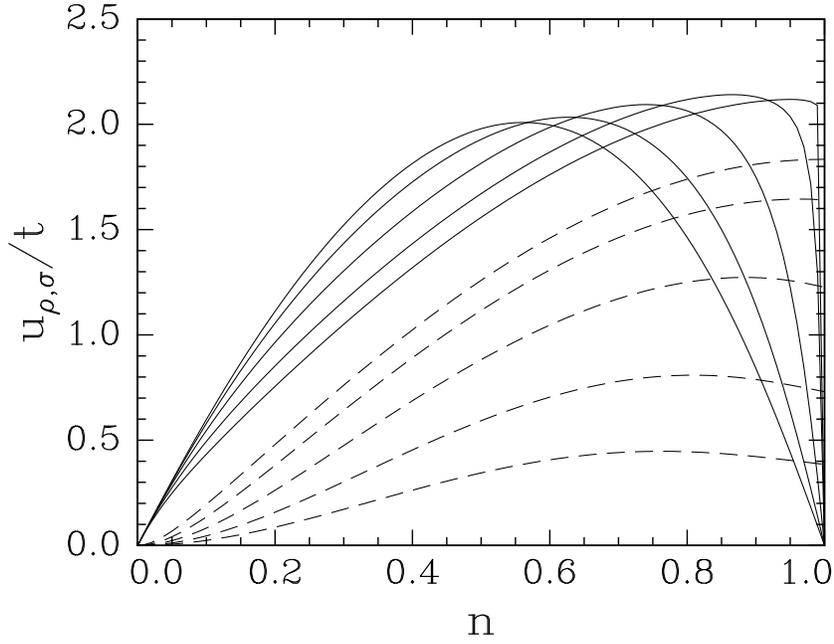

Figure 11: *The charge and spin velocities $u_\rho$ (full line) and $u_\sigma$ (dashed line) for the Hubbard model, as a function of the band filling for different values of $U/t$: for $u_\sigma$ $U/t = 1, 2, 4, 8, 16$ from top to bottom, for $u_\rho$ $U/t = 16, 8, 4, 2, 1$ from top to bottom in the left part of the figure.*

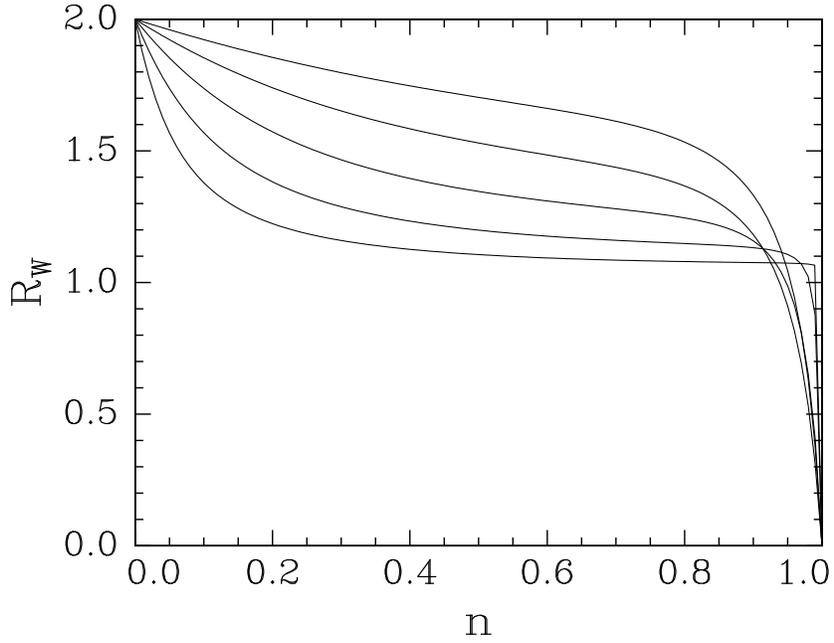

Figure 12: *The Wilson ratio $R_W$ for the one-dimensional Hubbard model, as a function of the band filling for different values of $U/t$ ($U/t = 16, 8, 4, 2, 1$ for the top to bottom curves).*

To obtain the parameter $K_\rho$ from the exact solution note that the gradient of the phase field $\phi_\rho$ is proportional to the particle density, and in particular a constant slope of $\phi_\rho$ represents a change of total particle number. Consequently, the coefficient $u_\rho/K_\rho$ in eq. (3.4) is proportional to the variation of the ground state energy $E_0$ with particle number:

$$\frac{1}{L}\frac{\partial^2 E_0(n)}{\partial n^2} = \frac{\pi}{2}\frac{u_\rho}{K_\rho} \ . \tag{4.37}$$

Note that this quantity is the inverse of the compressibility. Equation (4.37) now allows the direct determination of $K_\rho$: $E_0(n)$ can be obtained solving (numerically) Lieb and Wu's [Lieb and Wu, 1968] integral equation, and $u_\rho$ is already known. The results for $K_\rho$ as a function of particle density are shown in fig.13 for different values of $U/t$. For small $U$ one finds in all cases agreement with the perturbative expression, eq. (4.36), whereas for large $U$ $K_\rho \to 1/2$. The limiting behaviour for large $U$ can be understood noting that for $U = \infty$ the charge dynamics of the system can be described by noninteracting *spinless* fermions (the hard-core constraint then is satisfied by the Pauli principle) with $k_F$ replaced by $2k_F$. Consequently one finds a contribution proportional to $\cos(4k_F x)x^{-2}$ in the density-density correlation function, which from eq. (3.17) implies $K_\rho = 1/2$. One then finds an asymptotic decay like $\cos(2k_F x)x^{-3/2}\ln^{1/2}(x)$ for the spin-spin correlations, eq.(3.18), and an exponent $\alpha = 1/8$ in the momentum distribution function. The result $\alpha = 1/8$ has also been found by Anderson and Ren (preprint), and by Parola and Sorella [Parola and Sorella, 1990]. Ogata and Shiba's numerical results [Ogata and Shiba, 1990] are quite close to these exact values.

As is apparent from fig.13, the strong-coupling value $K_\rho = 1/2$ is also reached in the limits $n \to 0, 1$ *for any positive $U$*. For $n \to 0$ this behaviour is easily understood: the effective interaction parameter is $U/v_F$, but $v_F$ goes to zero in the low-density limit (corresponding to the diverging density of states). The limit $n \to 1$ is more subtle: in this case nearly every site is singly occupied, with a very low density of holes. The only important interaction then is the short range repulsion between holes, which can be approximated by treating the *holes* as a gas of spinless noninteracting fermions. Using (4.37), one then again finds $K_\rho = 1/2$.

We note that in the whole parameter region, as long as the interaction is repulsive one always has $K_\rho < 1$, which means that magnetic fluctuations are enhanced over the noninteracting case. On the other hand, superconducting pairing is always suppressed.

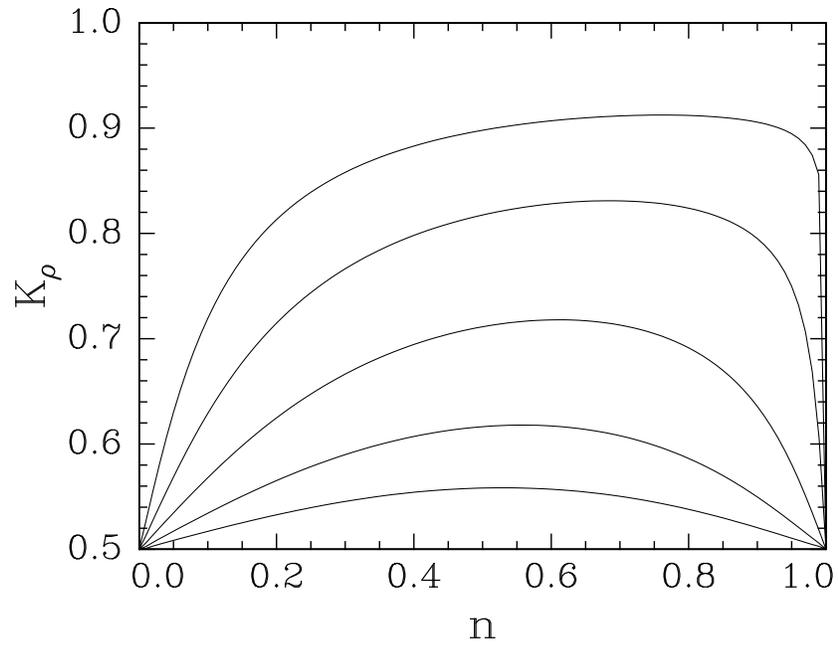

Figure 13: *The correlation exponent $K_\rho$ as a function of the bandfilling n for different values of U ($U/t = 1, 2, 4, 8, 16$ for the top to bottom curves). Note the rapid variation near $n = 1$ for small U.*

It should be emphasized here that the results of fig.13 are valid for $n \to 1$, but not for $n = 1$. In fact, in that latter case, there is a gap in the charge excitation spectrum, as expected from the umklapp term (3.7), and the correlations of $\phi_\rho$ become long ranged. Close to half–filling, the asymptotic behaviour of the charge part of correlation functions like (3.17) is essentially determined by the motion of the holes. Writing the density of holes as $\rho = 1/(1-n)$ one then expects a crossover of the form [Schulz, 1980]

$$\langle n(x)n(0) \rangle \approx \cos(2k_F x)[1 + (\rho x)^2]^{K_\rho/2} x^{-1} \ln^{-3/2}(x) \tag{4.38}$$

for the $2k_F$ part of the density correlation function, and similarly for other correlation functions. Clearly, only for $x \gg 1/\rho$ are the asymptotic power laws valid, whereas at intermediate distances $1 \ll x \ll 1/\rho$ one has effectively $K_\rho = 0$. Clearly, the form (4.38) provides a smooth crossover as $n \to 1$.

Results equivalent to the present ones can be obtained using the conformal invariance of the Hubbard model [Frahm and Korepin, 1990, Kawakami and Yang, 1990a]. These results have subsequently be generalized to the case with an applied magnetic field [Frahm and Korepin, 1991].

### 4.3.2 Transport properties

The exact solution of Lieb and Wu can also be combined with the long–wavelength effective Hamiltonian (3.3) to obtain some information on the frequency–dependent conductivity $\sigma(\omega)$. On the one hand, from eq. (3.13) there is a delta function peak at zero frequency of weight $2K_\rho u_\rho$. On the other hand, the total oscillator strength is proportional to the kinetic energy [Baeriswyl et al., 1986]:

$$\sigma_{tot} = \int_{-\infty}^{\infty} \sigma(\omega) d\omega = -\pi \langle H_{kin} \rangle / L \ . \tag{4.39}$$

Thus, both the weight of the dc peak and the relative weight of the dc peak in the total conductivity can be obtained and are plotted in fig.14. As expected, far from half–filling, all the weight in $\sigma_{tot}$ is in the dc peak. For exactly half–filling the dc conductivity vanishes, due to the existence of a gap $\Delta_c$ for charge excitations created by umklapp scattering, and all the weight is at $\omega > \Delta_c$. Fig.2 then shows that as $n \to 1$ umklapp scattering progressively transfers weight from zero to high frequency. The crossover is very sharp for small or large $U$, but rather smooth in intermediate cases ($U/t \approx 16$). This nonmonotonic behaviour as a function of $U$ can be understood noting that initially with increasing $U$ umklapp scattering plays an increasingly important role. Beyond $U/t \approx 16$, however, the spinless–fermion picture becomes more and more appropriate, and at $U = \infty$ one again has all the weight in the dc peak. The linear vanishing of $\sigma_0$ as $n \to 1$ implies a linear variation of the ratio $n/m^*$ with "doping".

An interesting question is the sign of the charge carriers, especially close to the metal–insulator transition. The standard way to determine this, the sign of the Hall constant, is useless in a one–dimensional system. As an alternative, the thermopower can be used which is negative (positive) for electron (hole) conduction. In general, calculation of the thermopower is a nontrivial task, as the curvature of the bands plays an important role, and the approximate form of the Hamiltonian (3.3) is therefore insufficient. Moreover, both charge ans spin entropies can play a role. However, close to the metal–insulator transition $u_\rho \ll u_\sigma$, and therefore the entropy of the charge degrees of freedom is much bigger than the spin entropy. In the presence of umklapp scattering, which becomes important close to half–filling, the charge part of the Hamiltonian can be transformed into a model of massive fermions, with energy–momentum relation $\varepsilon_k = \pm(v^2 k^2 + \Delta^2)^{1/2}$ [Emery et al., 1976]. $\Delta$ is the charge excitation gap created by umklapp scattering. In general, the quasi–particles interact, however close to half–filling this interaction can be eliminated [Schulz, 1980]. At half-filling all negative energy states are filled, all positive energy states are empty. Doping with a concentration $n^*$ of holes, some of the negative energy states become empty and only states with $|k| > k_F^* \propto n^*$ are filled. Because of the vanishing interaction, a standard formula for the thermopower can be used [Chaikin et al., 1976] and gives

$$S = \frac{\pi^2 k_B^2 T}{6|e|} \frac{\Delta^2}{v^2(k_F^*)^2(v^2(k_F^*)^2 + \Delta^2)^{1/2}} \ , \tag{4.40}$$

i.e. *approaching the metal–insulator transition from $n < 1$, the thermopower is hole–like*, whereas obviously far from the transition ($n \ll 1$) it is electron–like. The exactly opposite behaviour can be found for $n > 1$.

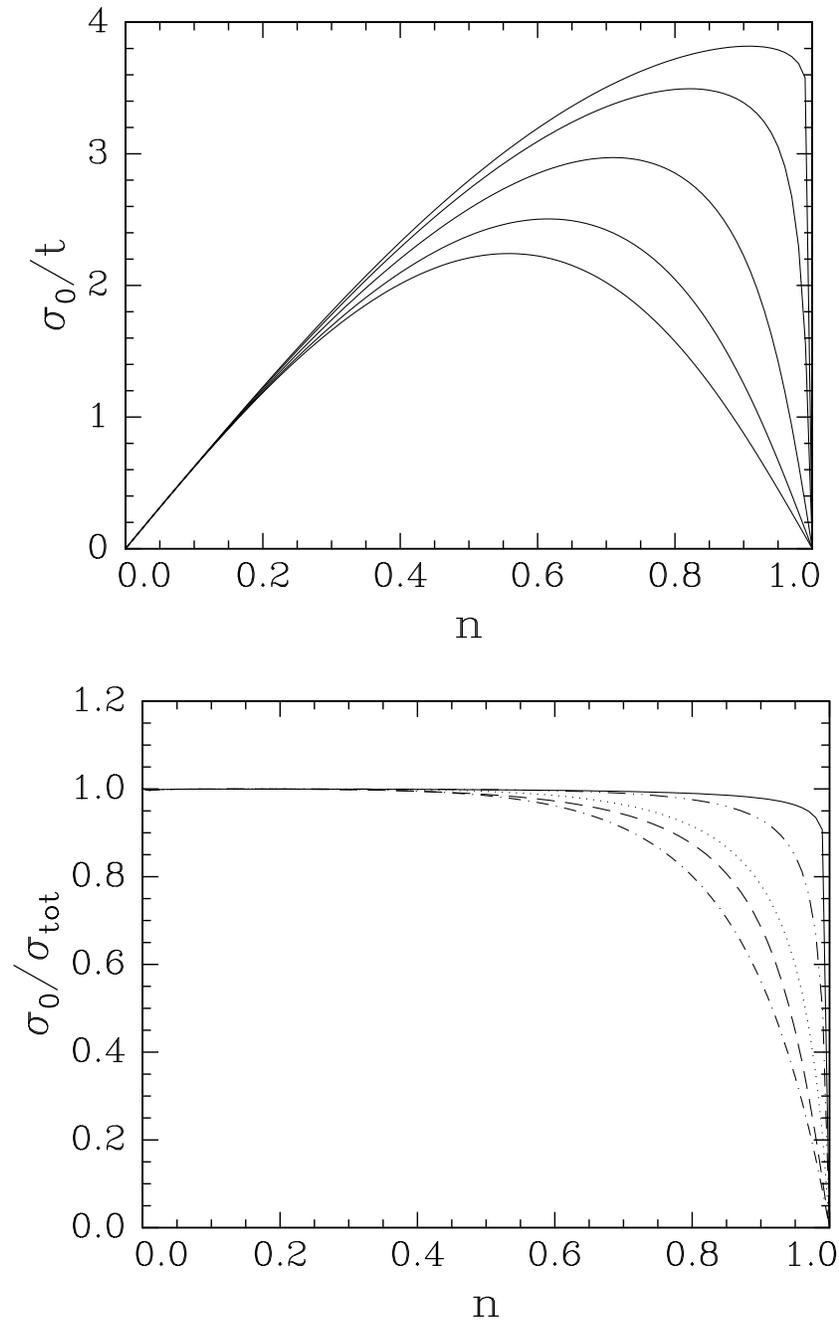

Figure 14: Top: *The weight of the dc peak in $\sigma(\omega)$ as a function of bandfilling for different values of $U/t$ ($U/t = 1, 2, 4, 8, 16$ for the top to bottom curves).*
Bottom: *Variation of the relative weight of the dc peak in the total conductivity oscillator strength as a function of the bandfilling n for different values of U: $U/t = 1$ (full line), 4 (dashed), 16 (dash–dotted), 64 (dotted), and 256 (dash–double-dotted).*

### 4.3.3 Spin–charge separation

The Hubbard model also provides a rather straightforward interpretation of the spin–charge separation discussed above. Consider a piece of a Hubbard chain with a half–filled band. Then for strong $U$ there will be no doubly–occupied sites, and because of the strong short–range antiferromagnetic order the typical local configuration will be

$$\cdots \uparrow\downarrow\uparrow\downarrow\uparrow\downarrow\uparrow\downarrow\uparrow\downarrow \cdots$$

Introducing a hole will lead to

$$\cdots \uparrow\downarrow\uparrow\downarrow\uparrow\ O\ \uparrow\downarrow\uparrow\downarrow\uparrow\downarrow \cdots$$

and after moving the hole one has (note that the kinetic term in the Hamiltonian does not flip spins)

$$\cdots \uparrow\downarrow\ O\ \uparrow\downarrow\uparrow\uparrow\downarrow\uparrow\downarrow\uparrow\downarrow \cdots$$

Now the hole is surrounded by one up and one down spin, whereas somewhere else there are two adjacent up spins. Finally, a few exchange spin processes lead to

$$\cdots \uparrow\downarrow\ O\ \uparrow\downarrow\uparrow\downarrow\uparrow\downarrow\uparrow\uparrow\downarrow \cdots$$

Note that the original configuration, a hole surrounded by *two* up spins has split into a hole surrounded by antiferromagnetically aligned spins ("holon") and a domain–wall like configuration, two adjacent up spins, which contain an excess spin 1/2 with respect to the initial antiferromagnet ("spinon"). The exact solution by Lieb and Wu contains two types of quantum numbers which can be associated with the dynamics of the spinons and holons, respectively. We thus notice that spinons and holons [Kivelson et al., 1987, Zou and Anderson, 1988] have a well-defined meaning in the present one–dimensional case.

The above pictures suggest that, as far as charge motion is concerned, the Hubbard model away from half–filling can be considered as a one–dimensional harmonic solid, the motion of the holes providing for an effective elastic coupling between adjacent electrons. This picture has been shown to lead to the correct long–distance correlation functions for spinless fermions [Haldane, 1981b, Emery, 1987]. For the case with spin, this suggests that one can consider the system as a harmonic solid with a spin at each site of the elastic lattice (lattice site = electron in this picture).

Let us now show that this gives indeed the correct spin correlation functions. In a continuum approximation, the spin density then becomes

$$\boldsymbol{\sigma}(x) = \sum_m \boldsymbol{S}_m \delta(x - x_m) \; , \tag{4.41}$$

where the sum is over all electrons. After a Fourier transformation of the delta function the spin–spin correlation function becomes

$$\langle \boldsymbol{\sigma}(x) \cdot \boldsymbol{\sigma}(0) \rangle = \frac{1}{(2\pi)^2} \int dq \, dq' \sum_{m,m'} e^{-iqx} \langle \boldsymbol{S}_m \cdot \boldsymbol{S}_{m'} e^{i(qx_m + q'x_{m'})} \rangle \; . \tag{4.42}$$

The exchange energy between adjacent spins is always antiferromagnetic, whether there is a hole between them or not, and consequently the low–energy spin dynamics always is that of an antiferromagnetic chain of localized spins. Under the additional assumption that the spin–spin correlations on an elastic lattice depend mainly on the average exchange constant and not so much on the fluctuations induced by motion of the electrons, the average in (4.42) factorizes into separate spin and charge factors. Following the hypothesis about harmonic motion of the electrons, we write $x_m = R_m + u_m$, where $R_m = m/|1-n|$ is the average position of the $m$th electron and $u_m$ the displacement with respect to this position. Note that the "harmonic solid hypothesis" implies that $u_{m+1} - u_m$ is small, but not necessarily $u_m$ and $u_{m+1}$ separately. In the averages over atomic positions now all terms with $q \neq q'$ vanish, and one has

$$\langle \boldsymbol{\sigma}(x) \cdot \boldsymbol{\sigma}(0) \rangle \approx \int dq \sum_{m,m'} e^{-iqx} \langle \boldsymbol{S}_m \cdot \boldsymbol{S}_{m'} \rangle e^{iq(R_m - R_{m'})} \langle e^{iq(u_m - u_{m'})} \rangle \; . \tag{4.43}$$

The average over $u_m$ in (4.43) has a power law behaviour:

$$\langle e^{iq(u_m - u_{m'})} \rangle \approx |m - m'|^{-\alpha(q)} \; ,$$

with $\alpha(q) \propto q^2$, i.e. it has a smooth $q$–dependence. On the other hand, the long–distance behaviour of the spin–spin correlations of an antiferromagnetic spin chain is [Luther and Peschel, 1975, Affleck et al., 1989]

$$\langle \mathbf{S}_m \cdot \mathbf{S}_{m'} \rangle \approx (-1)^{m-m'} |m - m'|^{-1} \ln^{1/2} |m - m'| \ .$$

Therefore in (4.43) the $q$–integration is dominated by terms with $q \approx \pi(1-n) = 2k_F$. Replacing the weakly $q$–dependent exponent $\alpha(q)$ by $\alpha(2k_F)$ one obtains

$$\langle \boldsymbol{\sigma}(x) \cdot \boldsymbol{\sigma}(0) \rangle \approx \int dq e^{-iqx} \sum_{m,m'} \langle \mathbf{S}_m \cdot \mathbf{S}_{m'} \rangle e^{iq(R_m - R_{m'})} |m - m'|^{\alpha(2k_F)} \quad (4.44)$$

$$= \cos(2k_F x) x^{-1-\alpha(2k_F)} \ln^{1/2}(x) \ . \quad (4.45)$$

With the identification $\alpha(2k_F) = K_\rho$, this is precisely the result (3.18). We thus have shown that the spin–spin correlations of a correlated electron system can in fact be understood as those of an elastic lattice of spins. In that picture, the motion of the holes then only provides the effective elasticity for the lattice.

### 4.3.4 The metal–insulator transition

It seems worthwhile here to compare the metal–insulator transition in the one–dimensional Hubbard model with other scenarios for strongly correlated fermion systems in higher dimension (see the review by Vollhardt [Vollhardt, 1984]). In the "nearly localized" picture, effective mass effects predominate and enhance both the specific heat and the spin susceptibility. Consequently, the Wilson ratio $(1/(1 + F_0^a)$ in Fermi liquid language) remains nonzero as the metal–insulator is approached. On the other hand, in the "nearly ferromagnetic" (or paramagnon) picture, only the spin susceptibility is enhanced significantly, and therefore $R_W$ can be much larger than unity. The behaviour found here in the one–dimensional case is quite different from both these scenarios: generally $R_W < 2$, and approaching the metal–insulator transition $R_W \to 0$. This occurs because generally an enhancement of the mass of the *charge* carriers (i.e. a decrease of $u_\rho$) has no influence on the spin degrees of freedom (see fig.1). This is rather straightforwardly understood in terms of *spin–charge decoupling*, as explained in the previous section: charge and spin excitations move nearly independently of each other, and in particular the spin dynamics is determined by antiferromagnetic nearest–neighbor exchange. In particular the spin susceptibility remains finite even when the mass of the charge carrier approaches infinity.

Let us discuss the metal–insulator transition in more detail. The fact that $u_\rho$ and $\sigma_0$ vanish linearly as $n \to 1$ seems to be consistent with a divergent effective mass at constant carrier density because $u_\rho \approx 1/m^*, \sigma_0 \approx n/m^*$. A constant carrier density is also consistent with the fact that $k_F = \pi n/2$ is independent of $U$. It is *not consistent* with the hole–like sign of the thermopower as $n \to 1$ from below, nor with the electron–like sign as $n \to 1$ from above: if the carriers are holes, the carrier density is the density of holes: $n^* = 1 - n$. Treating the holes as spinless fermions, as already mentioned before, one expects $\sigma_0 \to 0$ because $n^* \to 0$, and $\gamma \to \infty$ because the density of states of a one–dimensional band diverges at the band edges. This agrees with what was found explicitly in section 4.3.1. What is not so easily understood in this picture is the fact that $k_F$ (i.e. the location of the singularity of $n_k$) is given by its free–electron value $\pi n/2$, rather then being proportional to $n^*$. One should however notice that $n_k$ is given by the single–particle Green's function, which contains both charge and spin degrees of freedom. The location of $k_F$ then may possibly be explained by phase shifts due to holon–spinon interaction. This is in fact suggested by the structure of the wavefunction of the exact solution [Ogata and Shiba, 1990].

The magnetic properties do not agree with what one expects from an effective mass diverging as $n \to 1$: $u_\sigma$ and therefor $\chi$ remain finite. Moreover, the NMR relaxation rate would have the behaviour $1/T_1 = \alpha T + \beta\sqrt{T}$, where the first (Korringa) term comes from fluctuations with $q \approx 0$, whereas the second term comes from antiferromagnetic fluctuations with $q \approx 2k_F$. None of these properties is strongly influenced by the diverging effective mass observed e.g. in the specific heat. This fact is of course a manifestation of the separation between spin and charge degrees of freedom.

### 4.3.5 Other models

For more complicated models, e.g. the "extended Hubbard model"

$$H = -t \sum_{i,s}(a^\dagger_{is}a_{i+1,s} + a^\dagger_{i+1,s}a_{is}) + U \sum_i n_{i\uparrow}n_{i\downarrow} + V \sum_i n_i n_{i+1} \ , \tag{4.46}$$

exact eigenvalues can not be obtained in the thermodynamic limit. The parameters in eq. (4.37) can however be calculated reliably for finite systems, and this gives rather good results, as shown previously [Schulz, 1990b].

Exact exponents can be obtained for the model (4.46) in the limit $U \to \infty$: then one has effectively spinless fermions (with $k_F \to 2k_F$) with nearest neighbour interaction, a model which can be exactly solved using the Jordan–Wigner transformation into the XXZ spin chain. In particular, the $4k_F$–component of (3.17) is related to the correlation function of $S_z$. From the known results [Luther and Peschel, 1975] one obtains, for a quarter–filled band ($n = 1/2$), $K_\rho = 1/(2 + (4/\pi)\sin^{-1}(v))$, $u_\rho = \pi t\sqrt{1-v^2}/\cos^{-1}(v)$, with $v = V/2|t|$. Now $K_\rho < 1/2$ is possible. For $v > 1$ the system is in a dimerized insulating state. Approaching the insulating state from $v < 1$ both $K_\rho$ and $u_\rho$ remain finite, i.e. $\sigma_0$ jumps to zero at $v = 1$. For $n \neq 1/2$ the parameters $u_\rho, K_\rho$ can be obtained from numerical results [Haldane, 1980]. Quite generally, one has $K_\rho > 1/8$, but $K_\rho = 1/2$ for $n \to 0, 1$, independent of $v$. On the other hand, $u_\rho \to 0$ as $n \to 1/2$ for $v > 1$, i.e. in that case the weight of the dc conductivity goes to zero continuously, the point $(v, n) = (1, 1/2)$ is thus highly singular. The same type of singularity also occurs at $U = 0, n = 1$ in the Hubbard model. Interestingly enough, one has $K_\rho > 1$ if $V < -\sqrt{2}|t|$, i.e. a finite amount of

nearest–neighbor attraction is sufficient to lead to divergent superconducting fluctuations even for infinite on–site repulsion. Also note that the singularities in $u_\rho$ and $K_\rho$ at $v = -1$ (attractive interaction) represent a point of phase separation.

For the $t - J$ model, there is one exactly solvable point ($t = J$) where exact exponents can be found using the Bethe ansatz [Kawakami and Yang, 1990b]. Away from this point, eq. (4.37) has been used to obtain $K_\rho$ from numerical data [Ogata et al., 1991]. For large $J$ there is a phase with predominantly superconducting fluctuations. Exact exponents have also been found for a model of fermions with $1/r^2$ interaction [Kawakami and Yang, 1991].

# 5  Conclusion

In these notes, we have seen that using vastly different techniques, ranging from perturbation expansions via the bosonization method to exact solutions, one obtains a rather complete picture of many physical properties of interacting one–dimensional fermions. Probably the most important feature arising is the Luttinger liquid like behaviour, characterized by non–universal power laws, together with the separation of the charge and spin dynamics. One should also notice that there are no qualitative differences between weak and strong correlation.

Does this behaviour generalize to higher dimensions? In fact, Anderson has suggested that Luttinger liquid behaviour might also occur in two dimensions [Anderson, 1990], as well as in coupled chain systems [Anderson, 1991]. At least for the coupled–chain case, this is in contradiction with standard scaling [Schulz, 1991] and renormalization group arguments [Bourbonnais and Caron, 1991, Fabrizio et al., 1992].

In the weak–coupling limit and close to half–filling, the two–dimensional Hubbard model has some similarities with the one–dimensional case: one has a similar, albeit more complicated problem of coupled particle–particle and particle–hole singularities [Dzyaloshinskii, 1987, Schulz, 1987]. Upon doping, the antiferromagnetic structure becomes an incommensurate spin–density wave [Schulz, 1990c]. However, quite unlike the one–dimensional case, upon further doping antiferromagnetism vanishes alltogether. Moreover, strong correlation seems to be quite different from the weakly correlated case: for a slightly doped antiferromagnet, both a spiral state [Shraiman and Siggia, 1989] and phase separation [Emery et al., 1990] have been proposed, but a linearly polarized antiferromagnet seems to be excluded. It thus seems that the simple and direct connection between weak and strong correlation of the one–dimensional case does not simply carry over to two (or higher) dimensions.

**Acknowledgement** I'm grateful to colleagues in Orsay, in particular T. Giamarchi and D. Jérome, for many stimulating discussions on the subject of these notes. This work was in part supported by ESPRIT basic research contract no. 3121.

# References

Affleck, I., Gepner, D., Ziman, T., and Schulz, H. J. (1989). *J. Phys. A*, 22:511.

Anderson, P. W. (1970). *J. Phys. C*, 3:2346.

Anderson, P. W. (1987). *Science*, 235:1196.

Anderson, P. W. (1988). *Frontiers and Borderlines in Many Particle Physics*. North Holland, Amsterdam.

Anderson, P. W. (1990). *Phys. Rev. Lett.*, 64:1839.

Anderson, P. W. (1991). *Phys. Rev. Lett.*, 67:3844.

Anderson, P. W., Yuval, G., and Hamann, D. R. (1970). *Phys. Rev. B*, 1:4464.

Andrei, N. (1988). lecture notes, Theoretical Advanced Study Institute, Brown University.

Baeriswyl, D., Carmelo, J., and Luther, A. (1986). *Phys. Rev. B*, 33:7247.

Baxter, R. J. (1982). *Exactly Solved Models of Statistical Mechanics*. Academic Press.

Bethe, H. A. (1931). *Z. Phys.*, 71:205.

Bourbonnais, C. and Caron, L. G. (1991). *Int. J. Mod. Phys. B*, 5:1033.

Bourbonnais, C., Creuzet, F., Jérome, D., Bechgaard, K., and Moradpour, A. (1984). *J. Phys. (Paris) Lett.*, 45:L755.

Bychkov, Y. A., Gorkov, L. P., and Dzyaloshinskii, I. E. (1966). *Sov. Phys. JETP*, 23:489.

Chaikin, P. M., Greene, R. L., Etemad, S., and Engler, E. (1976). *Phys. Rev. B*, 13:1627.

Coll, C. F. (1974). *Phys. Rev. B*, 9:2150.

Dzyaloshinskii, I. (1987). *Sov. Phys. JETP*, 66:848.

Emery, V. J. (1979). *Highly Conducting One-Dimensional Solids*, page 327. Plenum, New York.

Emery, V. J. (1987). *Low-Dimensional Conductors and Superconductors*, page 47. Plenum, New York.

Emery, V. J. (1992). this volume.

Emery, V. J., Kivelson, S. A., and Lin, H. Q. (1990). *Phys. Rev. Lett.*, 64:475.

Emery, V. J., Luther, A., and Peschel, I. (1976). *Phys. Rev. B*, 13:1272.

Essler, F. H. L., Korepin, V. E., and Schoutens, K. (1991). *Phys. Rev. Lett.*, 67:3848.

Fabrizio, M., Parola, A., and Tosatti, E. (1992). Trieste preprint.

Faddeev, L. D. and Takhtajan, L. A. (1981). *Phys. Lett. A*, 85:375.

Frahm, H. and Korepin, V. E. (1990). *Phys. Rev. B*, 42:10553.

Frahm, H. and Korepin, V. E. (1991). *Phys. Rev. B*, 43:5663.

Gaudin, M. (1967). *Phys. Lett. A*, 24:55.

Georges, A. and Kotliar, G. (1992). *Phys. Rev. B*, 45:6479.

Giamarchi, T. and Schulz, H. J. (1988). *Phys. Rev. B*, 37:325.

Giamarchi, T. and Schulz, H. J. (1989). *Phys. Rev. B*, 39:4620.

Haldane, F. D. M. (1980). *Phys. Rev. Lett.*, 45:1358.

Haldane, F. D. M. (1981a). *J. Phys. C*, 14:2585.

Haldane, F. D. M. (1981b). *Phys. Rev. Lett.*, 47:1840.
Heeger, A. J., Kivelson, S., Schrieffer, J. R., and Su, W. P. (1988). *Rev. Mod. Phys.*, 60:781.
Heidenreich, R., Seiler, R., and Uhlenbrock, D. A. (1980). *J. Stat. Phys.*, 22:27.
Jérome, D. and Schulz, H. J. (1982). *Adv. Phys.*, 31:299.
Kawakami, N. and Yang, S. K. (1990a). *Phys. Lett. A*, 148:359.
Kawakami, N. and Yang, S. K. (1990b). *Phys. Rev. Lett.*, 65:2309.
Kawakami, N. and Yang, S. K. (1991). *Phys. Rev. Lett.*, 67:2493.
Kivelson, S., Rokhsar, D., and Sethna, J. (1987). *Phys. Rev. B*, 35:8865.
Lieb, E. H. and Wu, F. Y. (1968). *Phys. Rev. Lett.*, 20:1445.
Luther, A. and Emery, V. J. (1974). *Phys. Rev. Lett.*, 33:589.
Luther, A. and Peschel, I. (1974). *Phys. Rev. B*, 9:2911.
Luther, A. and Peschel, I. (1975). *Phys. Rev. B*, 12:3908.
Luttinger, J. M. (1963). *J. Math. Phys.*, 4:1154.
Mattis, D. C. (1974). *J. Math. Phys.*, 15:609.
Mattis, D. C. and Lieb, E. H. (1965). *J. Math. Phys.*, 6:304.
Ogata, M., Luchini, M. U., Sorella, S., and Assaad, F. (1991). *Phys. Rev. Lett.*, 66:2388.
Ogata, M. and Shiba, H. (1990). *Phys. Rev. B*, 41:2326.
Parola, A. and Sorella, S. (1990). *Phys. Rev. Lett.*, 64:1831.
Pernici, M. (1990). *Europhys. Lett.*, 12:75.
Pouget, J. P. (1988). *Highly Conducting Quasi-One-Dimensional Crystal Semiconductors and Semimetals*, volume 27, page 87. Pergamon, New York.
Schulz, H. J. (1980). *Phys. Rev. B*, 22:5274.
Schulz, H. J. (1987). *Europhys. Lett.*, 4:609.
Schulz, H. J. (1990a). *Phys. Rev. Lett.*, 65:2462.
Schulz, H. J. (1990b). *Phys. Rev. Lett.*, 64:2831.
Schulz, H. J. (1990c). *Phys. Rev. Lett.*, 64:1445.
Schulz, H. J. (1991). *Int. J. Mod. Phys. B*, 5:57.
Shiba, H. (1972). *Phys. Rev. B*, 6:930.
Shraiman, B. I. and Siggia, E. D. (1989). *Phys. Rev. Lett.*, 62:1564.
Sólyom, J. (1979). *Adv. Phys.*, 28:209.
Thacker, H. (1982). *Rev. Mod. Phys.*, 53:253.
Voit, J. and Schulz, H. J. (1988). *Phys. Rev. B*, 37:10068.
Vollhardt, D. (1984). *Rev. Mod. Phys.*, 56:99.
Vollhardt, D. (1992). this volume.
Woynarovich, F. (1982). *J. Phys. C*, 15:85.
Woynarovich, F. (1983). *J. Phys. C*, 16:5293.
Yang, C. N. (1967). *Phys. Rev. Lett.*, 19:1312.
Zhang, S. (1990). *Phys. Rev. Lett.*, 65:120.
Zou, Z. and Anderson, P. W. (1988). *Phys. Rev. B*, 37:627.